\documentclass[twocolumn,floats,pra,showpacs]{revtex4-1}

\usepackage{amsmath}
\usepackage{amsfonts}

\usepackage{tikz}

\newcommand {\be}{\begin{equation}}
\newcommand {\ee} {\end{equation}}
\newcommand {\bea}{\begin{eqnarray}}
\newcommand {\eea} {\end{eqnarray}}

\newcommand{\bk}{{\bf k}}

\newcommand{\bx}{{\bf x}}

\newcommand{\hilb}{\mathcal{H}}

\newcommand{\ket}[1]{\left| #1 \right>}
\newcommand{\bra}[1]{\left< #1 \right|}

\newcommand{\kket}[1]{\left| #1  \right>}

\begin{document}


\title{Tensor network trial states for chiral topological phases in two dimensions\\
and a no-go theorem in any dimension}
\author{J. Dubail and N. Read}
\affiliation{Department of Physics, Yale
University, P.O. Box 208120, New Haven, CT 06520-8120, USA}
\date{October 23, 2015}

\begin{abstract}
Trial wavefunctions that can be represented by summing over locally-coupled degrees of freedom
are called tensor network states (TNSs); they have seemed difficult to construct for two-dimensional
topological phases that possess protected gapless edge excitations. We show it can be done for
chiral states of free fermions, using a Gaussian Grassmann integral, yielding $p_x \pm i p_y$
and Chern insulator states, in the sense that the fermionic excitations live in a topologically
non-trivial bundle of the required type. We prove that any strictly short-range quadratic parent
Hamiltonian for these states is gapless; the proof holds for a class of systems in any dimension
of space. The proof also shows, quite generally, that sets of compactly-supported
Wannier-type functions do not exist for band structures
in this class. We construct further examples of TNSs that are analogs of fractional (including non-Abelian)
quantum Hall phases; it is not known whether parent Hamiltonians for these are also gapless.
\end{abstract}

\pacs{pacs} 

\maketitle

\section{Introduction}

Our theoretical understanding of quantum phases of matter frequently relies on the use of
trial ground state wavefunctions whose properties serve as the paradigm for an entire phase, such as
the Bardeen-Cooper-Schrieffer (BCS) wavefunction in the theory of superconductivity \cite{BCS}, and
the Laughlin wavefunction in the fractional quantum Hall effect \cite{laugh}; both examples
have generalizations that describe distinct phases.
In recent work, there has been progress in understanding the structure of ground states of
generic short-range lattice Hamiltonians, especially for ``topological'' phases. A Hamiltonian can be
defined as being in a topological phase if it possesses a gap in the
bulk energy spectrum above the ground state energy. For many ground states, a representation as
a {\em tensor network state} (TNS), in which the amplitude of a basis configuration of the local
degrees of freedom is expressed as a product of tensors that involve auxiliary degrees of freedom
which are then summed over, can be found. In our definition of a TNS, we further impose that the
auxiliary degrees of freedom live in finite-dimensional local Hilbert spaces, with only short-range
couplings; when they live on the links of a lattice in $d$ space dimensions and couple only to the
physical degrees of freedom on the end of that link, the state is known as a {\em projected entangled pair
state} (PEPS) \cite{PEPS}, or in the one-dimensional case as a {\it matrix product state} (MPS)
\cite{Cirac_MPS}. TNSs are also sometimes referred to as ``tensor product states'', but we avoid this
term because it both risks confusion with product states that have no entanglement, and conflicts with
mathematical terminology for tensor product spaces.

In one dimension it is known that MPSs provide enough variational freedom to approximate the ground
state in any topological phase \cite{Hastings_MPS}. In more than one dimension, it is known that many
trial states and exact ground states of special lattice Hamiltonians that lie in a
topological phase that does not exhibit topologically-protected (e.g.\ chiral) gapless edge
excitations can be viewed as TNSs, and it seems natural to expect the approximation results from
one dimension to generalize. But for chiral topological phases in two dimensions,
such as quantum Hall states and $p\pm ip$ BCS paired states, it has seemed difficult to obtain
a TNS in the phase, even as a trial state. Most attempts did not impose locality \cite{fermionPEPS}, while
B\'eri and Cooper \cite{CooperBeri}, by truncating a flat-band Hamiltonian, obtained some local
tensor networks that approximate expectation values in certain trial states.

In this paper we exhibit some fairly simple trial TNSs that belong to chiral topological phases
in two dimensions, in the sense that the states exhibit the desired topological structure. We begin
by constructing explicit examples of translationally-invariant Gaussian
(i.e.\ free-) fermionic TNSs, in each of which the TNS is defined locally by a few tensors. The basic
examples are TNS versions of a topologically-nontrivial $p\pm ip$ BCS paired state of fermions \cite{RG},
and of a filled band with Chern number equal to one \cite{thou_hald}. In each of these, there is
a Hamiltonian with the TNS as ground state and a gapped but dispersionless (``flat-band'')
fermion excitation-energy spectrum; these Hamiltonians have power-law-decaying matrix elements
in position space \cite{cirac}. Moreover, a consequence of the ground state being a free-fermion TNS is
that there are single-fermion destruction operators that annihilate the TNS, and are strictly
short-range, in the sense that their support is {\em compact} (i.e.\ a bounded region). Using these,
there are many ways to construct a ``parent'' Hamiltonian that has uniformly bounded support for
its terms, each of which annihilates the TNS.
The constructions used in these examples generalize to other free-fermion phases,
including other symmetry classes, and to higher dimensions. [For a one-dimensional lattice, they yield
the matrix-product ground state of the Kitaev chain \cite{KitaevChain}.]
We also calculate the entanglement spectrum of some of our two-dimensional examples, which exhibits some
unusual features.

However, for the states in free-fermion topological phases with chiral edge excitations in two
dimensions that
we construct, we find that the parent Hamiltonian is always {\em gapless} in the bulk. (In addition,
these TNSs always have power-law, rather than the desired exponential, correlations for some local
operators, as noted in some examples in Ref.\ \cite{cirac}.) We prove a general No-Go Theorem, which
states that for any free-fermion TNS in
a class of topological phases (those in which the vector bundle formed by the filled bands is
topologically non-trivial as a complex vector bundle, ignoring additional symmetries; these include the
chiral examples above) in {\em any} space dimension $d$, any short-range single-particle parent
Hamiltonian always has a gapless bulk energy spectrum. (Such a TNS also has power-law correlations in
the bulk.)  A gap can be
produced for the given TNS only by using a Hamiltonian with long-range matrix elements.
A slight variation of the same argument also proves, for the same class of free-fermion band structures,
that a set of compactly-supported single-particle states that, when Fourier transformed, span the states
in the filled band at every point $\bk$ in the Brillouin zone (``Wannier-type'' states), cannot exist
unless the filled-band bundle is topologically trivial when viewed as a complex vector bundle,
ignoring symmetries.

Back in two dimensions, we go further by utilizing the free-fermion TNSs and (similarly to earlier authors
\cite{fermionPEPS,CooperBeri}) imposing local constraints on a system of several copies of a
TNS, producing further TNSs. We argue that such constructions produce the chiral topological
phases associated with a variety of Chern-Simons theories \cite{witten} or fractional quantum Hall states,
including non-Abelian topological phases \cite{MR}. We do not at present have parent Hamiltonians for
these TNSs, though these should exist based on general principles of TNSs; we expect that they would be
gapless.

The paper is organized as follows: in Sec.\ \ref{sec:tns} we describe our construction of free-fermion TNSs
and their properties. In Sec.\ \ref{sec:nogo} we state and prove the No-Go Theorem. In Sec.\
\ref{sec:nonfree}, we briefly describe the constructions for non-free chiral topological phases in two
dimensions. Some Appendices explain some additional points or give examples.

\section{Gaussian tensor network states for topological phases}
\label{sec:tns}

In this Section we describe the basic constructions in the two-dimensional case. We begin with generalities
on Gaussian fermionic TNSs, and continue with examples of topologically non-trivial phases. Then we
discuss the annihilation operators that (in real space) have compact support, and the parent Hamiltonians
(which are quadratic in fermion operators) for which the TNS is the exact ground state. There is one set
of Hamiltonians that each has ``flat-band'' energy spectrum, but is not short range, while there is another
set each of which is short-range but gapless. We also show that the entanglement spectrum of these TNSs
has a peculiar form.

\subsection{Gaussian fermionic TNS}

The square lattice $\mathbb{Z}^2 \subset \mathbb{R}^2$ is generated by the
two vectors ${\bf i}= (1,0)$ and ${\bf j}=(0,1)$. Our physical degrees of freedom are fermions, with $n$
orbitals per site; the creation/annihilation modes obey the canonical anti-commutation relations $\{
c^\dagger_{{\bf x},\alpha} , c_{{\bf x}',\alpha'} \} \, = \, \delta_{\alpha,\alpha'}\, \delta_{{\bf x},
{\bf x}'}$. The fermion vacuum $\kket{0}$ is annihilated by all the $c_{{\bf x},\alpha}$. A class of
translation-invariant Gaussian TNSs is constructed as follows: to every site ${\bf x}$, we associate a
set of (real) Grassmann variables $\xi^p_{{\bf x}}$, $p \in \left\{1,\dots,P \right\}$. To every edge
$\left<{\bf x}' {\bf x} \right>$ (${\bf x}' \in \{ {\bf x}+{\bf i}, {\bf x} + {\bf j} \}$), we associate
the expression
\begin{equation}
	\label{eq:exp1}
	e^{\sum_{p,q}\xi^p_{{\bf x}+{\bf i}} \,  A^h_{pq} \, \xi^q_{{\bf x}}} \quad {\rm or  } \quad  e^{\sum_{p,q}\xi^p_{{\bf
x}+{\bf j}} \,  A^v_{pq} \, \xi^q_{{\bf x}}}
\end{equation}
with $A^h_{pq}, A^v_{pq} \in \mathbb{C}$. The superscripts $h$, $v$ stand for ``horizontal''
and ``vertical'';
in what follows, we will leave these superscripts implicit. To every site ${\bf x}$, we associate a weight
and a generating function of physical particles onsite:
\begin{equation}
	\label{eq:exp2}
	e^{\sum_{p,q}\xi^p_{\bf x} \,  B_{pq}  \, \xi^q_{\bf x}} \times e^{\sum_{q,\alpha}\xi^q_{\bf x}
 \kappa^\alpha_q c_{{\bf
x},\alpha}^\dagger},
\end{equation}
with $B_{pq} = - B_{qp} \in \mathbb{C}$ and $\kappa^\alpha_q\in\mathbb{C}$. Grassmann variables
anticommute with the physical creation/annihilation operators; thus, the two exponentials in
(\ref{eq:exp1}) and those in (\ref{eq:exp2}) all commute, both onsite and at different sites. After
taking the product, we integrate out the Grassmann variables to obtain a translation-invariant Gaussian
TNS:
\begin{equation}
	\label{eq:Gaussian}
	\kket{\psi} \, \propto \, \int [d\xi]  \, \prod_{{\rm edges} \, \left<{\bf x } {\bf y} \right>}
e^{\xi^t_{{\bf x}} \cdot A \cdot \xi_{{\bf y}}} \prod_{{\rm sites}\,{\bf z}}\, e^{\xi^t_{{\bf z}} \cdot
B \cdot \xi_{{\bf z}}} e^{\xi^t_{{\bf z}} \cdot \kappa \cdot c^\dagger_{{\bf z}}} \kket{0}.
\end{equation}
Here we have used the compact notation $\xi_{{\bf x}} = (\xi^1_{{\bf x}}, \dots , \xi^P_{{\bf x}})$,
$c^\dagger_{{\bf x}} = (c^\dagger_{{\bf x},1}, \dots , c^\dagger_{{\bf x},n})$, and matrices $A$, $B$,
$\kappa$. In (\ref{eq:Gaussian}),
some ordering of the Grassmann variables must be chosen to define the ``measure'' $\int [d\xi]$. By
construction, the state $\kket{\psi}$ is always a free-fermion BCS paired state.
It is possible to write the state (\ref{eq:Gaussian}) in a form that is closer to the usual form of
TNS \cite{PEPS}, with the Grassmann variables living on the edges
rather than on the sites; see Appendix \ref{app:peps}.

\subsection{Example in $p_x-ip_y$ phase}
\label{p-ip}

Our first example has $n=1$ orbital per site; we
use $P=2$ Grassmann variables on each site. The $A$, $B$, $\kappa$-matrices are:
\begin{subequations}
\begin{eqnarray}
	\xi_{{\bf x}+{\bf i}}^t \cdot A^h \cdot \xi_{{\bf x}} & = & \left( \begin{array}{cc} \xi^1_{{\bf x}
+{\bf i}} & \xi^2_{{\bf x}+{\bf i}} \end{array} \right) \left(\begin{array}{cc} -i & \lambda   \\
-\lambda & -i \end{array} \right) \left( \begin{array}{c} \xi^1_{{\bf x}} \\ \xi^2_{{\bf x}} \end{array}
 \right), \quad \\
	\xi_{{\bf x}+{\bf j}}^t \cdot A^v \cdot \xi_{{\bf x}} & = & \left( \begin{array}{cc} \xi^1_{{\bf x}+
{\bf j}} & \xi^2_{{\bf x}+{\bf j}} \end{array} \right) \left(\begin{array}{cc} 1 & \lambda   \\  -\lambda
& -1 \end{array} \right) \left( \begin{array}{c} \xi^1_{{\bf x}} \\ \xi^2_{{\bf x}} \end{array} \right),
 \quad \\
	\xi_{{\bf x}}^t \cdot B \cdot \xi_{{\bf x}} & = & \left( \begin{array}{cc} \xi^1_{{\bf x}} &
\xi^2_{{\bf x}} \end{array} \right) \left(\begin{array}{cc} 0 & -2\lambda  \\ 2\lambda  & 0
\end{array} \right) \left( \begin{array}{c} \xi^1_{{\bf x}} \\ \xi^2_{{\bf x}} \end{array} \right),
\\
	\xi_{{\bf x}}^t \cdot \kappa  \cdot c^\dagger_{{\bf x}} & = & \left( \begin{array}{cc} \xi^1_{{\bf x}}
& \xi^2_{{\bf x}} \end{array} \right) \left(\begin{array}{c}  \kappa_1 \\ 0 \end{array} \right) \left(
\begin{array}{c} c^\dagger_{{\bf x}} \end{array} \right) .
\end{eqnarray}
\end{subequations}
$\lambda \in \mathbb{R}$ and $\kappa_1 \in \mathbb{C}$ are two variational parameters.

With these
matrices, (\ref{eq:Gaussian}) gives a state $\kket{\psi_D}$; its behavior is easily
analyzed in momentum space. The Fourier modes of the particle creation operator are defined by
$c^\dagger_{{\bf x}} \, = \, \int \frac{d^2 {\bf k}}{(2\pi)^2} e^{-i {\bf k} \cdot {\bf x}} ~
c^\dagger_{{\bf k}}
$,
where the integral is over the first Brillouin zone $\left[-\pi,\pi \right]^2$. Similarly, for the
Grassmann variables
$
\xi_{{\bf x}} \, = \, \int \frac{d^2 {\bf k}}{(2\pi)^2} e^{-i {\bf k} \cdot {\bf x}} \xi_{\bf k}
$.
The terms $\prod_{ {\bf z}} e^{\xi^t_{{\bf z}} \cdot \kappa \cdot c^\dagger_{{\bf z}}}$ and $e^S \,\equiv
\, \prod_{\left<{\bf x} {\bf y} \right>} e^{\xi^t_{{\bf x}}\cdot A \cdot \xi_{{\bf y}} }  \prod_{{\bf z}}
e^{\xi^t_{{\bf z}} \cdot B \cdot \xi_{{\bf z}} }$ in (\ref{eq:Gaussian}) become respectively:
\begin{subequations}
\begin{eqnarray}
	&& \exp \left( {\int \frac{d^2 {\bf k}}{(2\pi)^2}} \xi^t_{-{{\bf k}}} \cdot  \kappa \cdot
c^\dagger_{{\bf k}}  \right), \\
	& & \exp \left[ \int \frac{d^2 {\bf k}}{(2\pi)^2} \left( \begin{array}{cc} \xi^1_{-{\bf k}}  &
\xi^2_{-{\bf k}} \end{array} \right) S_{\bf k} \left( \begin{array}{c} \xi^1_{{\bf k}}  \\ \xi^2_{{\bf k}}
\end{array} \right) \right],
\end{eqnarray}
\end{subequations}
where $S_{\bf k}$ is the $2\times 2$ matrix
\begin{equation}
	\label{eq:Sk}
	 \left( \begin{array}{cc} \sin k_x + i \sin k_y & -\lambda\left(2 - \cos k_x  - \cos k_y\right)  \\
  \lambda\left(2 - \cos k_x - \cos k_y  \right)  & \sin k_x - i \sin k_y  \end{array} \right).
\end{equation}
The integral over all the Fourier modes of the Grassmann variables is easily performed. It yields the
familiar BCS form (see e.g.\ Ref. \cite{RG})
\begin{equation}
	\label{eq:BCSmomentum}
	\kket{\psi_D} \, \propto \, \exp \left( \frac{1}{2}\int \frac{d^2{\bf k}}{(2\pi)^2} \,  g_{{\bf k}} \,
c^\dagger_{{\bf k}} c^\dagger_{-{\bf k}} \right) \kket{0},
\end{equation}
with a {\em pairing function\/} $g_{\bf k}$ which is the $(1,1)$ matrix element of the inverse matrix of
$S_{{\bf k}}$, and so is a component of the propagator of the Grassmann variables:
\begin{equation}
	\frac{g_{\bf k}}{\kappa_1^2} \, = \, \frac{\int \left[ d \xi \right] ~e^S ~ \xi^1_{\bf k} \xi^1_{-
{\bf k}}}{\int \left[ d \xi \right] ~e^S} \, = \, \left[ S_{{\bf k}}^{-1} \right]_{11}.
\end{equation}
Explicitly,
\begin{equation}
	\label{eq:gk}
	\frac{g_{{\bf k}}}{\kappa_1^2} \, = \, \frac{\sin k_x - i \sin k_y}{\left(\sin k_x\right)^2 +
\left( \sin k_y \right)^2 + \lambda^2 \left[ 2 -\cos k_x - \cos k_y  \right]^2},
\end{equation}
and $g_{-\bk}=-g_\bk$ for all $\bk$. As $\bk\to{\bf 0}$, $g_\bk$ diverges as
$g_\bk \sim \kappa_1^2/(k_x+ik_y)$, and is non-diverging at other $\bk$. Hence, in position space,
$g({\bf x})\sim \kappa_1^2/(x+iy)$ as $|{\bf x}|\to\infty$, where ${\bf x} \equiv{\bf x}_i
-{\bf x}_j$ represents
the separation of the members $i$, $j$ of a pair. These properties are sufficient to show the state is
in the non-trivial $p-ip$ phase \cite{RG} in symmetry class D; they can also be related to a Chern number
in $\bk$-space (see below). One can show that the average density of particles scales as
$\sim|\kappa_1|^4\ln (1/|\kappa_1|)$ as $\kappa_1\to 0$.

\subsection{Annihilation operators and parent Hamiltonians}

A state of the Gaussian (or BCS) form in
eq.\ (\ref{eq:BCSmomentum}) is annihilated by (unnormalized) ``destruction'' mode operators
$c_\bk-g_\bk c_{-\bk}^\dagger$ for all $\bk$.
In our example, $g_\bk=v_\bk/u_\bk$ is clearly a ratio of two ``trigonometric polynomials'' (TPs;
i.e.\ polynomials
in $\sin k_x$, $\cos k_x$, $\sin k_y$, and $\cos k_y$) $u_\bk$ and $v_\bk$, which we assume have no common
TP factor other than a constant times an integer power of $e^{ik_x}$, times another of
$e^{ik_y}$. Then we define
\be
d_\bk = u_\bk c_\bk - v_\bk c_{-\bk}^\dagger.
\label{eq:dkop}
\ee
If we normalize the destruction operators (\ref{eq:dkop}) as
\be
\widehat{d}_\bk = \widehat{u}_\bk c_\bk- \widehat{v}_\bk c_{-\bk}^\dagger,
\ee
where $\widehat{u}_\bk=u_\bk/\sqrt{|u_\bk|^2+|v_\bk|^2}$,
$\widehat{v}_\bk=v_\bk/\sqrt{|u_\bk|^2+|v_\bk|^2}$, then $\widehat{d}_\bk$, $\widehat{d}_\bk^\dagger$
obey canonical anticommutation relations, $\{\widehat{d}_\bk,\widehat{d}_{\bk'}^\dagger\}=
(2\pi)^2\delta(\bk-\bk')$. For the ``flat-band'' Hamiltonian
\be
\widehat{H}_D=\int \frac{d^2\bk}{(2\pi)^2} \widehat{d}_\bk^\dagger\widehat{d}_\bk,
\ee
the fermion excitations $\widehat{d}_\bk^\dagger|\psi_D\rangle$ have nonzero and $\bk$-independent
energy for all $\bk$. Expanded in $c_\bk$, $c_\bk^\dagger$, the coefficients in $\widehat{H}_D$
are ratios of TPs, and are not real analytic in $k_x$, $k_y$
at $\bk=0$, but are elsewhere in the Brillouin zone. Hence in position space, $\widehat{H}_D$
contains terms that decay as powers of distance \cite{cirac}.

On the other hand, the operators $d_\bk$ contain coefficients that are TPs, and
so the inverse Fourier transform gives operators $d_{\bf x}$ that annihilate the TNS and are truly
local---they have compact support that surrounds $\bf x$. The existence of such operators is not
an accident. Any TNS has by
construction the property that if the sites of the system are bipartitioned into two sets, $A$ and $B$,
and $A$ is finite, then the rank of the Schmidt decomposition of the TNS can be bounded by
some constant
to the power of the surface area or perimeter of the region $A$. For free fermions, the reduced density
matrix again has the form of the exponential of a free-fermion Hamiltonian, and so for a free-fermion TNS
the number of fermion modes that can appear in the entanglement Hamiltonian is some constant times
the surface area, and so much smaller than the volume of region $A$ in general. It follows that
for such regions $A$, there must be linear combinations of fermion operators,
supported in region $A$, that annihilate the reduced density matrix, when acting on it from the left.
The operators $d_{\bf x}$ with support in region $A$ are a basis set for these operators in our case.
We note that these operators anticommute with one another, but do not in
general anticommute with the operators $d_{{\bf x}'}^\dagger$ at ${\bf x}'\neq{\bf x}$, except when
their supports are disjoint. By contrast, the $\widehat{d}_{\bf x}$ operators obey canonical
anticommutation relations and are true Wannier functions, but are not supported locally near
$\bf x$, instead they have long power-law tails due to non-analytic behavior in $\bk$-space at $\bk=0$
\cite{wannier}.

Using the operators $d_{\bf x}$, we can form other Hamiltonians that annihilate the TNS, for example:
\be
H_D=\int \frac{d^2\bk}{(2\pi)^2} d_\bk^\dagger d_\bk
=\sum_{\bf x} d_{\bf x}^\dagger d_{\bf x}.
\ee
This Hamiltonian is a sum of terms that have compact support, and each annihilates $|\psi_D\rangle$, so
it is a parent Hamiltonian, however from its $\bk$-space form we can see that it is gapless at
$\bk={\bf 0}$: by expressing it in terms of $\widehat{d}_\bk$, $\widehat{d}_\bk^\dagger$, we find
that the energy of a fermion excitation is $|u_\bk|^2+|v_\bk|^2$, which is $\propto \bk^2$ near $\bk=0$.

\begin{widetext}

\begin{figure}
	\includegraphics[width=\textwidth]{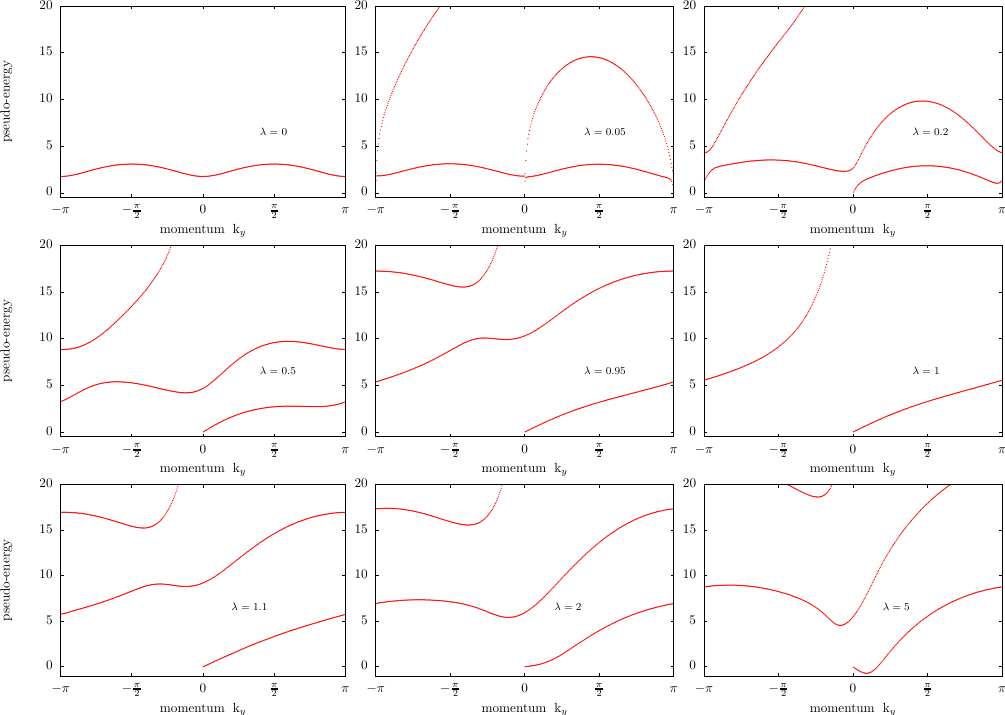}
	\caption{Single-particle entanglement spectrum of the $p-i \,p$ state $\ket{\psi_D}$ on an infinite
cylinder of circumference $L=200$. We take anti-periodic boundary conditions for the fermions around the
cylinder, such that $k_y \in \frac{2\pi}{L} (\mathbb{Z} + \frac{1}{2})$. Here $|\kappa_1| = 1$, and we
vary the parameter $\lambda \geq 0$.}
	\label{fig:sup_ES}
\end{figure}

\begin{figure}
	\includegraphics[width=\textwidth]{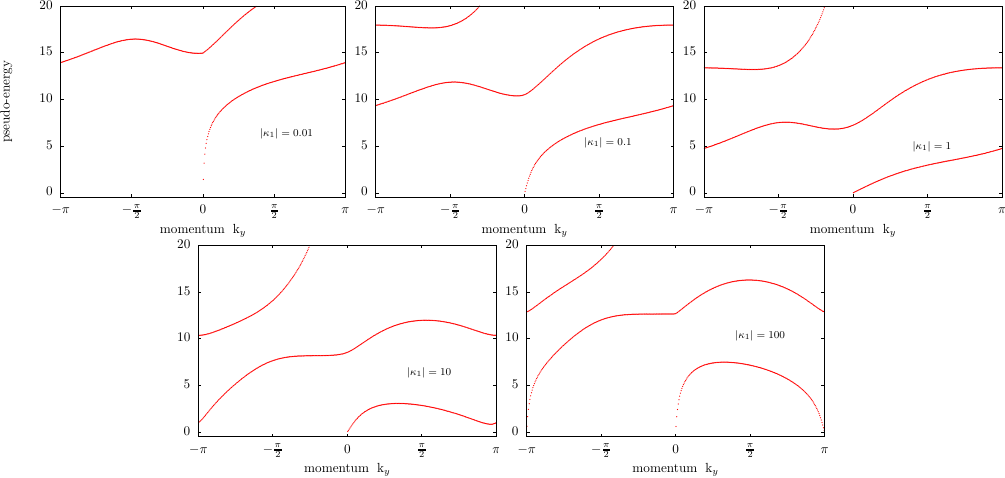}
	\caption{Same as in Fig.~\ref{fig:sup_ES}, with $\lambda$ fixed ($\lambda = 0.8$). We vary
$|\kappa_1|$.}
	\label{fig:sup_ES2}
\end{figure}

\end{widetext}

\subsection{Chern band example}

The preceding construction can be generalized to include more orbitals,
more singularities in $g_\bk$, more dimensions, or more symmetry (e.g.\ time reversal); we describe
one more example.
We take two copies of the previous example, by using $n=2$ orbitals per site; we view particles
occupying either type of orbital as two distinct types of particles, that have opposite charges under a
U(1) symmetry. In the TNS construction we use $P=4$ Grassmann variables, or $P=2$ {\em complex} Grassmann
variables, with the matrices constructed to respect the U(1) charge mentioned. Then we can obtain
(see Appendix \ref{app:chern} for more details) a TNS that is a BCS state with pairing only between
opposite particle
types,
\begin{equation}
	\kket{\psi_A} \, \propto \, \exp \left( \int \frac{d^2 {\bf k}}{(2\pi)^2} \, g^{21}_{{\bf k}}
\, c_{{\bf k},2}^\dagger c_{-{\bf k},1}^\dagger  \right) \kket{0},
\end{equation}
which conserves the U(1) symmetry; here $g_\bk^{21}$ is the same as $g_\bk$ above.
If we now perform a particle-hole transformation on the type-1 fermions, and so that
$c_{\bk,1}^\dagger\to c_{-\bk,1}$, and write $|0\rangle$ as $|1,0\rangle$, which is annihilated by all
$c_{{\bf x},1}^\dagger$ and $c_{{\bf x},2}$, we arrive at a particle-number conserving state
$|\widetilde{\psi}_A\rangle$ that represents a filled band (symmetry class A).  There are now two
types of operator $d_{\bk,\tilde{1}}$, $d_{\bk,\tilde{2}}$ that annihilate $\kket{\widetilde{
\psi}_A}$;
one of these is
\be
d_{-\bk,\tilde{1}}=u_\bk c_{\bk,1}^\dagger + v_\bk c_{\bk,2}^\dagger,
\label{dk1}
\ee
which is a creation operator associated with states in the filled band.
The state $|1,0\rangle$ (i.e.\ for $g_\bk=0$ for all $\bk$) describes topologically-trivial bands.
For general $g_\bk$ (which here does not have to be an odd function of $\bk$), the filled band
is non-trivial when its Chern number is nonzero; the other band has opposite Chern number
\cite{thou_hald}. The Chern number can be obtained (up to a choice of sign convention)
from the (generically isolated) points at which $g_\bk$ diverges, as the sum of the winding numbers of
each; the winding number can be defined as the winding of $g_\bk/|g_\bk|$ as $\bk$ traverses a
small circle about the point of divergence in the counterclockwise direction. Thus in our case, the
Chern number is 1. There are Hamiltonians $H_A$ ($\widehat{H}_A$) with similar properties as $H_D$
($\widehat{H}_D$) above: $H_A$ is short-range but gapless, while $\widehat{H}_A$ has flat-band form and
is gapped, but has power-law decaying matrix elements.

\subsection{Entanglement spectrum}

For chiral phases of matter, one expects the entanglement spectrum to exhibit gapless chiral modes
\cite{LiHaldane}. We have checked this for some of the Gaussian TNSs constructed earlier. Here we
illustrate this with our $p-i\,p$ example, which is a translation-invariant BCS state $\ket{\psi_D}$
defined by its pairing function
\begin{equation}
	g_{{\bf k}} \, = \, \frac{(\kappa_1)^2 (\sin k_x - i\, \sin k_y)}{ (\sin k_x)^2+(\sin k_y)^2
+ \lambda^2 [2- \cos k_x - \cos k_y]^2}.
\end{equation}
We put this Gaussian TNS on an infinite cylinder $(x,y) \in \mathbb{Z} \times [0,L]$, and chose the
bipartition $A \cup B$, $B  = \{(x,y) \,| \, x < 0  \}$ and $A = \{ (x,y) \, | \, x\geq 0 \}$. Notice
that the bipartition is translation-invariant in the $y$-direction. Since we are dealing with a Gaussian
state $\ket{\psi_D}$, its reduced density matrix $\rho_A \, = \, {\rm Tr}_B\, \ket{\psi_D} \bra{\psi_D}$
is the exponential of a sum of fermion bilinears. In other words, $- \ln \rho_A$ is a quadratic operator.
The entanglement spectrum ({\it i.e.} the spectrum of $-\ln \rho_A$) is a free fermion spectrum,
generated by
a set of single-particle {\it pseudo-energies} $\varepsilon_j$. Finally, $\rho_A$ (and thus $-\ln
\rho_A$) commutes with $T_y$, the generator of translations $(x,y) \mapsto (x,y+1)$. Thus, every
single-particle pseudo-energy is associated to a fixed momentum sector $k_y$. We plot the
single-particle entanglement spectrum $\varepsilon_p(k_y)$, which in general has more than one branch
(hence the subscript $p$).

By construction, the number of branches in the single-particle entanglement spectrum must be bounded by
the number of Grassmann variables per site that we use to define our TNS. Here there are two Grassmann
variables per site, so there are at most two branches in the single-particle spectrum (see
Figs.~\ref{fig:sup_ES} and \ref{fig:sup_ES2}). Since $\ket{\psi_D}$ only depends on $\lambda^2$, we can
focus on $\lambda
\geq 0$. For generic values of $\lambda$, there are two branches in the single-particle spectrum, but
when $\lambda = 0$ or $\lambda =1$, one of the two branches disappears (it goes to infinity, meaning
that the rank of $\rho_A$ is smaller for these values of $\lambda$), and one is left with a single branch.
The reduction of the rank can also be traced back to the properties of the pairing function $g_{{\bf k}}$
\cite{DR}. Indeed, $g_{{\bf k}}$ may be viewed as a rational function of the variable $e^{i k_x}$, and
has generically four simple poles, but has only two when $\lambda = 0$
or $\lambda=1$. This modifies the form of the Fourier transform $g_{k_y}(x) \, =\,\int
\frac{d k_x}{2\pi} e^{i k_x x} g_{{\bf k}}$, which determines the single-particle entanglement spectrum
\cite{DR}.

We clearly observe that, as long as $\lambda \neq 0$, there is one chiral gapless edge mode starting at
$k_y =0$, as expected. At $\lambda = 0$, this branch disappears; this is also expected, since at
$\lambda = 0$ the degree of the mapping ${\bf k} \mapsto g_{{\bf k}}$ suddenly changes, and the state
$\ket{\psi_D}$ does not belong to a chiral topological phase anymore.

\section{No-Go theorem}
\label{sec:nogo}

In this Section, we explain and prove the general No-Go Theorem for translation-invariant free-fermion
TNSs. First, we explain
various generalities about bands, bundles, and their parametrization using a Grassmannian manifold, then
introduce relevant notions of analytic and polynomial bundles. Then
we state and prove the No-Go Theorem, starting with special cases before giving the proof in the general
case. We also state a converse, that a gapped parent Hamiltonian exists in topologically-trivial cases,
and end with some remarks. Finally, we explain a variation of the proof that shows that complete sets
of compactly-supported Wannier functions cannot exist unless the filled-band bundle is topologically
trivial.

\subsection{Generalities on bands and bundles}

We begin by generalizing the construction in the Chern-band example.
In general, we consider a TNS for a system with $n$ orbitals per site
on a $d$-dimensional lattice, with $m\leq n$ filled bands (as in class A; those for class D map onto
these on taking two copies). In fact, we will begin with more general statements (occasionally indicating
how to specialize them for TNSs), before turning to the No-Go Theorem. The general free-fermion ground
state in this symmetry class has the form
\begin{equation}
	 \exp \left( \int \frac{d^d {\bf k}}{(2\pi)^d} \, \sum_{\alpha,\overline{\alpha}}g_{{\bf
k},\alpha\overline{\alpha}}
\, c_{{\bf k},\overline{\alpha}}^\dagger c_{{\bf k},\alpha}  \right) \kket{11\cdots,00\cdots0}.
\label{genTNS}
\end{equation}
Here $\alpha=1$, \ldots, $m$, $\overline{\alpha}=m+1$, \ldots, $n$, and the reference state
$\kket{11\cdots,00\cdots0}$ is annihilated by $c_{\bk,\alpha}^\dagger$ and $c_{\bk,\overline{\alpha}}$
for all $\bk$, or equivalently by $c_{\bx,\alpha}^\dagger$ and $c_{\bx,\overline{\alpha}}$ for all $\bx$.
We write $g_{\bk,\alpha\overline{\alpha}}$ as $g_{\bf k}$, which is an $m\times(n-m)$ matrix of functions
of $\bk$ in the Brillouin zone, say $[-\pi,\pi]^d$ for the hypercubic lattice; thus $\bk$ is always real,
except where explicitly stated otherwise. (In the case of a TNS, the entries of $g_\bk$ are ratios of TPs;
conversely, in Appendix \ref{app:alt} we show that such a form of $g_\bk$ can always be obtained from a
TNS, possibly one of more general form).

It will be useful later
if we point out here some arbitrariness in the specification of the state, in that the choice of which
orbitals are filled and which are empty on every site in the reference state was arbitrary.
We could have chosen a different reference state, in particular by replacing $\kket{11\cdots,00\cdots0}$
with a state annihilated by a different set of $m$ creation operators and by the annihilation operators
for the remaining $n-m$ orbitals orthogonal to these, for every site (the same set for each site $\bx$);
these sets of creation operators could be any linear combination of those used above. To preserve
the canonical anticommutation relations, the transformation on the set of all $n$ orbitals at each $\bx$
should be unitary; further, as it acts locally, it is independent of $\bk$ in $\bk$-space. In order that
the TNS be unchanged, the effect of such a transformation on $g_\bk$ has to be a fractional linear
transformation of the form
\be
g_\bk\to g_\bk'=(A+g_\bk B)^{-1}(C+g_\bk D)
\label{transfn}
\ee
for matrices $A$, $B$, $C$, $D$ of appropriate sizes, such that
\be
{\cal U}=\left(\begin{array}{cc}A&C\\B&D\end{array}\right)
\ee
is the $\bk$-independent $n\times n$ unitary matrix. [The reason why the transformation of $g_\bk$ takes
this form
may become clearer when we discuss the Grassmannian manifold $G(m,n)$ below.] We know that at some points
in $\bk$ space, say at $\bk=\bk_0$, $g_\bk$ may diverge; we can make $g_\bk$ finite at any given $\bk_0$,
using such a transformation if necessary. (Whether the resulting $g_\bk$ actually exists as a continuous
function at $\bk=\bk_0$ depends on the properties of the function $g_\bk$ with which we started; we discuss
this issue below.)

Generally (not only for TNSs), any one-fermion operator that annihilates the general free-fermion ground
state must take the form
\be
d_{-\bk}=\sum_{\alpha} u_{\bk,\alpha}
c^\dagger_{\bk,\alpha}+\sum_{\overline{\alpha}}v_{\bk,\overline{\alpha}}
c^\dagger_{\bk,\overline{\alpha}},\label{eq:d-k}
\ee
for the filled bands, or a similar form for the empty bands. At present, these operators can be
considered for any given $\bk$, without regard to how they depend on $\bk$. Here $u_\bk$ and $v_\bk$ are a
solution to the equation
\be
v_\bk=u_\bk g_\bk
\label{v=ug}
\ee
where $u_\bk$ is an $1 \times m$ matrix (or $m$-component row vector) and $v_\bk$ is an $1\times (n-m)$
matrix (or $n-m$-component row vector), for the given $\bk$, wherever it makes sense, that is wherever
$g_\bk$ is finite. Where $g_\bk$ is not finite,  we can make a transformation as above
to render it finite; this acts as a unitary change of basis on the $n$-dimensional vectors $(u,v)$.
The set of all solutions (viewed basis-independently) to this equation for all $\bk$ forms a space
consisting of a vector space of dimension at least $m$ for each $\bk$. (These vectors correspond to the
complex conjugates of the single-particle states in the filled band at that $\bk$.)

When $g_\bk$ is not too singular (we will explain the precise statement of this), this space has
the structure of a vector bundle \cite{milstash}; the base space of the bundle is the Brillouin zone, and
the fiber at each $\bk$ is the space of solutions to the equation at that $\bk$. To qualify as a bundle,
it must be possible to find, for any $\bk_0$, a neighborhood of $\bk_0$ in which there is a set of $m$
solutions which are linearly independent for each $\bk$ in the neighborhood, and each of which vary
continuously with $\bk$. (The dimension of the fiber as a complex vector space is also called the
{\em rank} of the bundle.) We refer to this bundle as the filled-band bundle. Our key assumption (other
than specializing to TNSs) is that $g_\bk$ is such that the corresponding single-particle
states in the filled bands define an $m$-dimensional vector bundle over the Brillouin
zone \cite{milstash}; otherwise the ground state does not lie in a topological phase. This means precisely
that in the neighborhood of any $\bk_0$, it is possible to transform so that $g_\bk$ is finite and
continuous (as a function of $\bk$) in that neighborhood. (As an example of the opposite situation, in
the situation when the ground state is associated with an energy spectrum that has a Dirac point, the
space of solutions does not form a bundle---$g_\bk$ itself, while finite, is not continuous at the
Dirac point, and so cannot even be uniquely defined there by continuity; see for example Ref.\ \cite{RG}.
This behavior is associated with the filled band ``touching'', or becoming degenerate with, an ``unfilled''
band at the Dirac point.)

Under these conditions, the states in the unfilled bands form an empty-band bundle, which is
complementary to the filled-band bundle, in the sense that the orthogonal direct sum of the two, which
consists of the full $n$-dimensional space of states at each $\bk$, is a bundle of
rank $n$. This rank-$n$ bundle is topologically trivial because it arises from the tight-binding model.
(We can also describe this by saying that both the filled and empty-band bundles are embedded as
sub-bundles in the trivial rank-$n$ bundle.) The (single-particle) Hilbert space of the
tight-binding model is a product of local single-particle Hilbert spaces in real space, and so in
$\bk$-space the states can be spanned by a set of $\bk$-independent vectors. We recall that a bundle of
rank $m$ is topologically trivial (as a complex vector bundle, with no reference to additional symmetries)
if and only if there is a set of $m$ sections that are linearly independent at each and every
$\bk$ value (it follows that these sections are non-vanishing everywhere); here and elsewhere a
{\em section} is a choice of a vector in the bundle at each $\bk$, and so is defined for every $\bk$ in
the base space (here, the Brillouin zone), and is further required to vary continuously with $\bk$ for
all $\bk$. For the case of $d=2$ dimensions, a non-zero Chern class (or number) implies that the bundle
is non-trivial \cite{milstash}. When the filled-band bundle is non-trivial, so is
that of the empty bands, so that their direct sum is trivial \cite{milstash}.

More explicitly, the condition that we have a well-defined filled-band bundle can be expressed by saying
that, in some neighborhood of any $\bk_0$, and possibly after a transformation, $g_\bk$ is finite,
and can be written in terms of a set of $m$ solutions $(u_\bk,v_\bk)$ which we assemble into matrices,
so that
\be
g_\bk=U_\bk^{-1}V_\bk,
\label{g=UinvV}
\ee
in the neighborhood, where $U_\bk$ is $m\times m$ and $V_\bk$ is $m\times (n-m)$ (the rows are
the vectors $u_\bk$, $v_\bk$). This is a consequence of the existence of linearly-independent sections
over the neighborhood of any point in the base space; by the choice of transformation, $U_\bk$ is
invertible throughout the neighborhood. It should go without saying that there may be no one choice of
transformation that makes these statements hold for all $\bk$.

The function
$g_\bk$, and the use of different coordinates related by a fractional linear transformation, can be
understood in terms of a Grassmannian manifold. The choice of an $m$-dimensional subspace in the (fixed)
$n$-dimensional space at each $\bk$ can be parametrized as corresponding to a point in the complex
Grassmannian
\be
G(m,n)=\frac{{\rm U}(n)}{{\rm U}(m)\times {\rm U}(n-m)}.
\ee
For $m=1$, $G(m,n)$ reduces to complex projective space ${\bf CP}^{n-1}$, and if also $n=2$, this becomes
${\bf CP}^1\cong S^2$, the 2-sphere. For the latter it is well known that if the vector is $(u,v)$ ($u$ and
$v$ complex numbers), then $g=v/u$ represents any point on the sphere as a point in the
plane by stereographic projection,  except for one pole which is mapped to infinity. By unitary rotation
in ${\rm U}(2)$ of the sphere, the pole omitted can
be mapped to any finite point (the transformation is fractional linear as above, with $A$, \ldots, $D$
replaced by complex numbers). The use of the matrices $U$, $V$, with $g=U^{-1}V$, and the fractional
linear transformations of $g$, generalize this to any $G(m,n)$. Thus a bundle of rank $m$ that is
continuously embedded as a sub-bundle of a trivial rank-$n$ bundle can be viewed as a continuous function
from the base space (the Brillouin zone) into the Grassmannian $G(m,n)$. Note that as we have described it,
our bundle is more than just a bundle as defined topologically \cite{milstash}: it has a given embedding
into the trivial $n$-dimensional bundle, and these are what are classified by the maps to the Grassmannian.

\subsection{Analytic bundles}

We introduce a further notion that does not hold for general bundles (or even for general sub-bundles),
and may be unconventional, but will be crucial for our discussion. A bundle of rank $m$ which is embedded
into a trivial bundle will be said to be {\em analytic} at a point $\bk_0$ in the base space if, in a
neighborhood of $\bk_0$ in the base space, there is a set of $m$ sections (defined over the neighborhood)
that are linearly independent and all of their components (relative to the given trivialization of the
rank $n$ bundle) are analytic functions of $\bk$, both properties holding in the neighborhood.
(Here analytic means as functions of the $d$ real variables $k_\mu$.) We define in a similar way the
notion of analytic in an open set in the base space. The term analytic, when used without
qualification
as to the neighborhood or open set in which it holds, means the bundle is analytic at all points in the
base space (for us, the Brillouin zone). We note that analyticity extends to
{\em complex} $\bk$ in a complex neighborhood of $\bk_0$ whenever it holds at a real $\bk_0$; this will be
important later (for the theory of analytic functions in several variables see e.g.\ Ref.\ \cite{gunning}).
We also note that the filled-band bundle is analytic if and only if the complementary empty-band bundle is.

The significance of analyticity is that if the embedded bundle (for example, the filled-band bundle)
is analytic, then a projection operator
onto the vectors in the fiber varies analytically with $\bk$ also. Thus for an analytic bundle,
taking the inverse Fourier transform of the projection operators onto the two bundles,
the resulting flat-band single-particle Hamiltonian will also be short range, at least in the
exponentially-decaying sense. Conversely, when the bands and
associated bundles are obtained from a short-range single-particle Hamiltonian that has an energy gap
between filled and empty bands at all $\bk$, the projection operators onto the filled and empty bands
are analytic (from basic results in matrix analysis on perturbations of eigenvalue problems);
this follows by taking the Fourier transform of a Hamiltonian whose ``hopping'' matrix elements decay
exponentially or faster with distance, so that in $\bk$-space the Hamiltonian is analytic in $\bk$.
Given an $n\times n$ projection operator $P_\bk$ onto the filled band, which is of rank $m$ and analytic
at all $\bk$, it is always possible to construct (e.g.\ by diagonalizing $P_\bk$) an $m\times n$ matrix
function $\phi_\bk$ in a neighborhood of any $\bk_0$, such that $\phi_\bk$ is analytic and
$P_\bk=\phi_\bk^\dagger\phi_\bk$
in the neighborhood. The rows of $\phi_\bk$ provide the local analytic sections, so the filled-band bundle
is analytic.

Like the definition of a bundle, the definition of an analytic bundle can be expressed in terms
of $g_\bk$ [after a suitable transformation of the form in (\ref{transfn})].
By definition of an embedded bundle, in some neighborhood of any $\bk$, say of $\bk_0$, the matrix
$W_\bk=(U_\bk,V_\bk)$ formed from the $m$ solutions in the neighborhood has non-vanishing minors of
rank $m$. Without loss of generality (that is, by
permuting the columns, which is a special case of a fractional linear transformation), we can assume
that $\det U\neq0$ at $\bk_0$. If the bundle is analytic in a neighborhood of $\bk=\bk_0$, then we can
assume $W_\bk$ is in fact analytic in the neighborhood, so $\det U\neq0$ also in a (possibly smaller)
neighborhood of $\bk_0$. Then it follows that $g_\bk$ is analytic in that neighborhood also. Conversely,
if (with a choice of a transformation) $g_\bk$ is analytic in a neighborhood of $\bk_0$ (and so does
not diverge in the neighborhood), then there are solutions $W_\bk$ that are analytic in that neighborhood
(for example, take $U_\bk=I$, $V_\bk=g_\bk$). Then if for all $\bk$ there is a transformation such that
$g_\bk$ is analytic at that $\bk$, then the bundle is analytic.

\subsection{TNSs give rise to polynomial bundles}

We now turn to the specific aspects that arise when we consider a TNS; in this Section, we return to
bundles that are not necessarily analytic. For a TNS, each entry of $g_\bk$ is
a ratio of TPs (i.e.\ trigonometric polynomials: see Sec.\ \ref{sec:tns}). Then it is natural to seek
solutions of eq.\ (\ref{v=ug}) such that $u_\bk$, $v_\bk$ also have TP entries. Each such solution is a
section of the filled-band bundle, and we call such a section a {\em TP section}. The TP
sections then give rise to the local (compact-support) operators in position space that annihilate the
ground state, via inverse Fourier transform of eq.\ (\ref{eq:d-k}). Of course, these do not give all the
sections in $\bk$-space of the filled-band bundle, either globally or over smaller regions. More general
sections can be obtained by taking linear combinations of the TP sections with arbitrary continuous
complex functions of $\bk$ as coefficients. Nonetheless, we will find that consideration of TP sections,
and of the simpler polynomial sections to be described next, will be very useful.

When considering solutions, or TP sections, say for the hypercubic lattice, it is
possible to work in terms of polynomials in $X_\mu=e^{ik_\mu}$ ($\mu=1$, \ldots, $d$) only, and not
their reciprocals (which would be allowed when working in TPs), because given a solution $u_\bk$, $v_\bk$
in TPs, we can multiply all components by positive powers of each $e^{ik_\mu}$ so that negative powers are
eliminated. (This trivial change in the solutions has the effect of translation in real space, which
is something we make use of anyway in obtaining operators with compact support.) Likewise $g_\bk$ itself
can be expressed in terms of ratios of such polynomials. Thus henceforth we will
be interested in solutions of eq.\ (\ref{v=ug}) with entries in the polynomial ring
\be
R={\bf C}[X_1,\ldots,X_d]
\ee
of such polynomials (and not in the TPs); we refer to these solutions simply as {\em polynomial sections}.
Solutions (with entries in $R$) to the eq.\ (\ref{v=ug}) are plentiful. We emphasize that polynomial
sections map to polynomial sections under basis changes that use constant unitary matrices
$\cal U$ as in a fractional linear transformation of $g_\bk$, eq.\ (\ref{transfn}) (a corresponding
statement holds for TP sections). $g_\bk$ itself can
be written in terms of some sets of $m$ solutions as in eq.\ (\ref{g=UinvV}), where now the entries of
$U_\bk$ and $V_\bk$ are in the ring $R$ and, excluding a trivial case in which $g_\bk$ is constant,
$U_\bk$ is invertible except possibly on a set of measure zero in the Brillouin zone.
Thus for example,
if the lowest common denominator (in $R$) of the entries of $g_\bk$ is $D_\bk$, we can write
$U_\bk=D_\bk I_m$, $V_\bk=U_\bk g_\bk$. (Note that $R$ is a unique factorization domain
\cite{reid,eisen1}, so that factorization can be used here. The lowest common denominator is unique up
to multiplication by an invertible element of $R$; the invertible elements are just the non-zero constants
in $\bf C$.) For lattices other than hypercubic, we can
similarly work in the ring of
polynomials in some combinations of exponentials of components of $\bk$. For example, for the triangular
lattice in $d=2$ dimensions with lattice spacing $1$, we can work with polynomials in $X_1=e^{ik_x}$ and
$X_2=e^{ik_x/2+\sqrt{3}ik_y/2}$. This works for any lattice, because the monomials in exponentials of
wavevectors that appear correspond to translations on the lattice, which form an abelian group, with $d$
generators. The indeterminates in the polynomials constituting the ring correspond to these generators.
Thus as a ring, $R$ always has the same structure for each $d$.

Similarly to eq.\ (\ref{v=ug}), the operators that annihilate the TNS by destroying states in the empty
band contain coefficients that are solutions of
\be
\widetilde{v}_\bk=g_\bk\widetilde{u}_\bk,
\ee
where $\widetilde{u}_\bk$ and $\widetilde{v}_\bk$ are column vectors of size $n-m$ and $m$ respectively,
and are now to be viewed as solutions in $R$. A set of $n-m$ of these that are linearly independent at
almost all values of $\bk$ can be assembled into matrices
$\widetilde{U}_\bk$, $\widetilde{V}_\bk$ of sizes $(n-m)\times(n-m)$ and $m\times(n-m)$, respectively,
so $g_\bk=\widetilde{V}_\bk\widetilde{U}_\bk^{-1}$. Finally, using the $m\times n$ matrix
$W=(U,V)$, and forming the $n\times(n-m)$ matrix $Z$ with the blocks
\be
\left(\begin{array}{c}\widetilde{V}\\-\widetilde{U}\end{array}\right),
\ee
(we will sometimes leave the $\bk$-dependence implicit, as here), the equations take the form
\be
WZ=0,
\label{WZ=0}
\ee
(linear equations with entries in $R$) which may either be solved to find solutions for
$w_\bk=(u_\bk,v_\bk)$ as rows of $W$, given any $Z$ of the form here for the given $g_\bk$, or
likewise for columns
\be
\left(\begin{array}{c}\widetilde{v}_\bk\\-\widetilde{u}_\bk\end{array}\right),
\ee
when $W$ is given. [These equations essentially assert that the states ($n$-component vectors) in the
filled band are orthogonal to those in the empty band at the same $\bk$.] In this form the equations
are manifestly invariant under a unitary transformation given by $W\to W {\cal U}$, $Z\to {\cal U}^{-1} Z$
for a constant unitary $n\times n$ matrix $\cal U$, the same one as in the transformation (\ref{transfn})
and following.

Finally, bundles that can be defined as a sub-bundle of a trivial bundle determined by solutions to eqs.\
(\ref{v=ug}) or (\ref{WZ=0})---i.e.\ linear
equations with polynomial coefficients---will be called {\em polynomial bundles}. Note that according to
our definitions, if there are points $\bk_0$ in the base space at which there are more than $m$
linearly-independent solutions for $w_\bk$ (over {\bf C}), then the bundle is defined by continuity at
those points; it was assumed that that is possible. It is also possible that solutions in $R$ span a space
of dimension smaller than $m$ at some $\bk_0$; again, since we obtain general sections by taking linear
combinations of those solutions using functions of $\bk$, we may be able to obtain continuous sections
that are linearly independent throughout a neighborhood of $\bk_0$, and it is our assumption that we can,
so the bundle is well-defined. (We will soon see, however, that in the context we wish to discuss, these
issues do not actually arise.)

\subsection{Statement and proof of No-Go Theorem}
\label{sec:nogo2}

In this Section we state and prove the No-Go Theorem for translation-invariant TNSs for free-fermion
systems.
We define the general TNS in $d$ dimensions to be of the general free-fermion form as in the present
Section \ref{sec:nogo}, with the additional condition that each entry of $g_\bk$ as a function of $\bk$
in the Brillouin zone is a ratio of polynomials taken from $R$. We are interested in parent Hamiltonians as
defined earlier; these are translation-invariant one-body Hamiltonians that annihilate the TNS, in which
the single-particle matrix elements are strictly short-range, that is they vanish unless the two
sites involved are separated by less than some constant. (We note that in class A, in which there is
a conserved
particle number, a chemical potential times particle number term can be added to the Hamiltonian. In our
definition, the chemical potential has already been fixed to lie between the filled and empty states. If
there is an energy gap between them, then the chemical potential can be changed without affecting the value
of the gap, or the ground state, as long as the chemical potential remains in the gap. Moving the chemical
potential out of this range would change the ground state and violate our assumptions.)

The No-Go Theorem arises from the confrontation between the properties of being analytic and being
polynomial. We have seen that if the Hamiltonian has a gap throughout the Brillouin zone, then the
filled-band bundle is analytic. At the same time, we have seen that for a TNS, the filled-band bundle is
a polynomial bundle (and that is the only consequence of having a TNS that we will use). We will ask if
the bundle is topologically non-trivial as a complex vector bundle. All of these properties have been
discussed in the immediately preceding sections. In view of the earlier discussion, the relevant Theorem
can now be stated as follows:

{\bf No-Go Theorem}: if a complex vector bundle is both polynomial and analytic, then it is
topologically trivial.

By contraposition, the
No-Go Theorem implies that for a TNS in which the filled-band bundle is nontrivial as a complex
bundle, there can be no parent Hamiltonian with a gap in its spectrum at the Fermi energy. Indeed, there
can be no gapped Hamiltonian even if its matrix elements are allowed to decay exponentially. We will also
show that conversely, when the filled-band bundle is analytic, there is a gapped parent Hamiltonian for any
TNS. We remark here that the mathematical statement and proof of the No-Go Theorem hold not only
for bundles over the Brillouin zone, which can be defined as the set $\{|X_\mu|=1:\mu=1,\ldots,d\}$ in
${\bf C}^d$, but also hold without modification for bundles with any domain in ${\bf C}^d$ as the base
space, in particular, with analyticity required throughout the domain. However, in the arguments, we treat
the case of the Brillouin zone, parametrized by $\bk$, which is real except when otherwise stated.

We will begin with a result that will be used repeatedly in the later proofs, and then give proofs of the
Theorem in a couple of special cases, as these are more easily understood, and
they build towards the general case. After stating some additional results that will be used, we will
proceed to the proof in the general case.

{\bf Proposition}: if $g_\bk$ has entries that are ratios of polynomials in $R$ and there is a $\bk_0$
such that $g_\bk$ is analytic in a neighborhood
of $\bk_0$, then the lowest common denominator $D_\bk$ of $g_\bk$ is non-vanishing at $\bk_0$, and
there is a set of $m$ solutions $w_\bk$ (with entries in $R$) to eq.\ (\ref{v=ug}) that are linearly
independent in the neighborhood of $\bk_0$.

We remark that the conclusion is similar to that in the discussion of analytic bundles above; the point
here is that $m$ {\em polynomial} sections $w_\bk$ can be used to span the bundle locally, in the
neighborhood of $\bk_0$.

{\bf Proof}: Analyticity at $\bk_0$ implies analyticity in some
{\em complex} neighborhood of $\bk_0$, that is at nearby complex values of $k_\mu$. If the denominator,
$D_\bk'$ say, in an entry in $g_\bk$ vanishes at $\bk_0$, then it does so on a variety, the solution set of
the polynomial equation $D_\bk'=0$. To ensure analyticity, the zero on the variety must be canceled not
only at $\bk_0$ but everywhere in the complex neighborhood; otherwise this entry of $g_\bk$
diverges at some complex $\bk$ points arbitrarily close to $\bk_0$ and so is not analytic at $\bk_0$
\cite{gunning}. For ratios of polynomials, this means that the numerator must contain a polynomial
factor that cancels those factors in the denominator that vanish in a neighborhood of $\bk_0$.  Hence
the lowest common denominator $D_\bk$ cannot vanish at $\bk_0$. We can obtain a matrix of
the form $W_\bk$ (with entries in $R$) using $U_\bk=D_\bk I_m$, and the rows of $U_\bk$ are linearly
independent in a neighborhood of $\bk_0$, because $\det U_\bk=D_\bk^m\neq0$ there. QED. We remark that
with this result, issues (mentioned above) of whether solutions to eq.\ (\ref{v=ug}) really define
a bundle do not arise once we assume that in some neighborhood of any $\bk_0$, $g_\bk$ is analytic after it
has been transformed to make it finite at $\bk_0$.

{\bf Corollary}: if a complex vector bundle of rank $m$ is both analytic and polynomial, then for any
$\bk$, there is a set of $m$ polynomial sections that are linearly independent in some neighborhood of
$\bk$. (This follows immediately from the Proposition.)

{\bf Proof} of the No-Go Theorem in the case $m=1$ (so the filled-band bundle has rank one): It is easy
to see that all solutions with entries in $R$ to eq.\ (\ref{v=ug}) (i.e.\ polynomial sections) are
multiples of that obtained by taking $u_\bk=D_\bk$ to be the lowest common denominator of the entries
of $g_\bk$ (in $R$). Call this basic solution $w_\bk$. By the Corollary, as the bundle is analytic,
for any $\bk=\bk_0$ there is a polynomial section that is non-vanishing in a neighborhood of $\bk_0$.
Since all polynomial sections are multiples of $w_\bk$, $w_\bk$ cannot vanish anywhere, which shows that
the bundle is trivial. QED.
As a second proof, note that the definition of $w_\bk$ implies that there can be no polynomial factor common
to all its entries, while analyticity implies that if $w_\bk$ vanished (as a vector) at some $\bk_0$,
then there would have to be such a common factor, yielding a contradiction.

{\bf Proof} of the No-Go Theorem for $m\geq 1$ under an additional hypothesis: namely,
we assume that all solutions to eq.\ (\ref{v=ug}) can be expressed as linear combinations
(with coefficients in $R$) of a set of $m$ basic solutions (or ``generators'') with entries in $R$;
we can write these polynomial sections as the matrix $W$. Then the argument is similar to the $m=1$ case.
By the Corollary, as the bundle is analytic, for any $\bk_0$, there is a set of $m$ polynomial sections
that are linearly independent in a neighborhood of $\bk_0$. As the generators (rows of $W$) are assumed
to span all solutions, they too must be linearly independent at $\bk_0$. As this holds for all $\bk$,
the set of generators gives the trivialization, and the bundle is trivial. QED.

Before turning to the proof in the general case, we first discuss the issues involved and the background
that it uses. The solutions to matrix equations like $wZ=0$ (where $w$ is an unknown row vector
and $Z$ is a known $n\times p$ matrix, for some $p$) in the polynomial ring $R$ may not all be expressible
as linear
combinations (over $R$) of a single set of $m$ solutions that are linearly independent over $R$. (This
basic fact about equations over a polynomial ring $R$ can be found in any text on commutative algebra,
such as Refs.\ \cite{reid,eisen1}.) Thus the additional hypothesis used in the preceding special case may
not hold.  Then, even though a particular set of $m$ solutions may be linearly
independent at some $\bk_0$ (as implied by the Proposition), that particular set might be linearly
dependent elsewhere, and there may not be one set of $m$ solutions that generates all such
locally-linearly-independent solution sets as linear combinations over $R$, so the argument breaks down.
Likewise, it is not clear either that there is a
set of $n-m$ column vectors $z$ that suffice to fully determine all the solutions $w$; it is possible that
more have to be used. The existence of such a set of size $n-m$ would again
lead to a simple proof as above (applied to the empty-band bundle). (The possible non-existence of such a
set is the reason we do not
assume that the form $WZ=0$ can be used, with $Z$ a known $n\times(n-m)$ matrix, but instead we continue
to work with $g$, which determines all the solutions. Note that when we stated above that there are matrices
with entries in $R$ such that $g=U^{-1}V$ or $g=\widetilde{V}\widetilde{U}^{-1}$, we never said in
the general case that those matrices
are unique, not even up to multiplication by invertible matrices.) This is the most challenging aspect of
the problem. To make further progress, we are forced to use more machinery.

The space of solutions to eq.\ (\ref{v=ug}) with entries in
$R$---that is, the space of polynomial sections---forms what is called a module over the ring $R$
(analogous to a vector space over a field): that is, the
space forms an Abelian group under addition, with the zero solution as the identity element, and any
element of the space times an element of $R$ yields another element of the space. Our module $M$ of
solutions to eq.\ (\ref{v=ug}) is a submodule of a ``free'' module of rank $n$ over $R$ (the
space of all $n$-component row vectors with entries in $R$), and so because $R$ is Noetherian (the Hilbert
Basis Theorem), $M$ has finite sets of generators, linear combinations (over $R$) of which span
the space \cite{reid,eisen1}. Given such a finite set of generators, we can form a matrix $\phi_1$, the
rows of which are the generators. However, unlike the case in which the ring is a field (and the module is
then a vector space), $M$ may not be free, that is (as already stated above) it may not be possible to find
a set of generators that are linearly independent (over $R$) \cite{reid,eisen1}. This implies that in this
case $\phi_1$ must have more than $m$ rows. Given a choice of a set of
generators, the linear relations (or syzygies) among them also form a module over
the same ring $R$, because the space of linear combinations of generators that equal zero obeys all the
conditions in the definition of a module. The module of relations in turn must be finitely generated but
may not be free, so a set of generators for it also obeys some relations, and so on. The Hilbert Syzygy
Theorem \cite{eisen1,eisen}, to be explained next, says that for the polynomial ring $R$ this so-called
chain of syzygies can be constructed so that it terminates in a bounded number of steps.

The Syzygy Theorem can be described in the following way \cite{eisen} (the application to our problem
is immediate): First, we note that a free module of any rank $n$ say, is isomorphic (over the polynomial
ring $R$) to the module consisting of all $n$-component row vectors
with entries in $R$. Then giving a set of generators for a
module $M$ is equivalent to specifying a map (a homomorphism of modules over $R$) of a free module onto $M$.
As our modules consist of spaces of row vectors with entries in $R$, this map $\phi_1$ can be described
as in the previous paragraph by a matrix that we also denote by $\phi_1$, whose rows are the generators.
Due to the generators not being linearly independent, the map of modules has a kernel (null space),
a subspace of the free module that is annihilated by $\phi_1$, and the kernel is a module which can itself
be described in a similar way, and so on. The {\bf Hilbert Syzygy Theorem} says that this construction can
be carried out so that there exists a so-called free resolution of the module (an exact sequence of
modules) of the form
\be
0\longrightarrow  F_d \stackrel{\phi_d} \longrightarrow F_{d-1} \stackrel{\phi_{d-1}}\longrightarrow
\cdots \stackrel{\phi_2}\longrightarrow F_1 \stackrel{\phi_1}\longrightarrow F_0
\label{syz_th}
\ee
in which $F_\mu$ ($\mu=1$, \ldots, $d$) is free and has rank $n_\mu$ over $R$, the rank of the free
module $F_0$ is $n_0=n$, and $0$ at the left end
is the trivial module. Each map $\phi_\mu$ can be viewed as an $n_\mu\times n_{\mu-1}$ matrix with entries
in $R$, and $\phi_{\mu+1}\phi_{\mu}=0$ for $\mu=1$, \ldots, $d$, with $\phi_{d+1}=0$.  (The case in which
the sequence is shorter can be thought of as $F_\mu=0$ for $\mu$ larger than some $\mu_0<d$.) The exactness
of the sequence means that the kernel of one map is precisely the image of its predecessor (as modules over
$R$). The module $M$ is the image of $\phi_1$, and also is by definition the kernel of an
additional map, corresponding to the equation $wZ=0$ for $w\in M$, so that the exact sequence can in
fact be extended one more step at the right. (It is in that form that the Syzygy Theorem is usually
stated.) The structure of such a free resolution of $M$ is described further by the
Buchsbaum-Eisenbud Theorem \cite{eisen}, which says that the rank (over $R$) of
each map $\phi_\mu$, which we write as ${\rm rank}_R\,\phi_\mu$, obeys the usual rank-nullity theorem of
linear algebra, which here reads
\be
{\rm rank}_R\, \phi_{\mu+1}+{\rm rank}_R\,\phi_\mu=n_\mu,
\ee
and also describes how the ranks of maps {\em evaluated at any particular complex $\bk$} behave: the rank
over ${\bf C}$ of the matrix of $\phi_\mu$ at some $\bk$, which we write as ${\rm rank}_{\bf
C}\,\phi_\mu(\bk)$, can be less than ${\rm rank}_R\,\phi_\mu$: ${\rm rank}_{\bf
C}\,\phi_\mu(\bk)\leq {\rm rank}_R\,\phi_\mu$ in general.

Even though in general the module of polynomial sections of the filled-band bundle is not free (as assumed
in the proof under that additional hypothesis), the Syzygy Theorem shows that the general case is not
much worse, because the sequence in eq.\ (\ref{syz_th}) terminates at the left. With that additional
information, we can now present a proof in the general case.

{\bf Proof} of the No-Go Theorem in the general case:
We first describe the filled-band vector bundle $E$. At each $\bk$, vectors in $E$ are solutions $w_\bk$
to eq.\ (\ref{v=ug}) (transformed to make $g_\bk$ finite, if necessary). As the bundle is analytic,
it follows from the Corollary that for any $\bk_0$ there is a set of solutions (with entries in $R$)
that are linearly independent in a neighborhood of $\bk_0$. Hence $\phi_1$ has ``full rank'' (namely $m$)
over {\bf C}, that is ${\rm rank}_{\bf C}\,\phi_1(\bk)={\rm rank}_R\,\phi_1=m$ at all $\bk$, and the
bundle $E$ can be
identified with the image of $\phi_1$, viewed as evaluated at each $\bk$ in the Brillouin zone.

Then we consider the sequence of vector bundles over the Brillouin zone
\be
0\longrightarrow  E_d \stackrel{\phi_d} \longrightarrow E_{d-1} \stackrel{\phi_{d-1}}\longrightarrow
\cdots \stackrel{\phi_2}\longrightarrow E_1 \stackrel{\phi_1}\longrightarrow E_0.
\ee
The bundles $E_\mu$ and $E_0$ are trivial bundles of the same ranks as the corresponding $F_\mu$ and $F_0$.
Maps of vector bundles are maps of the vector space (fiber) over each $\bk$ in one bundle into that at the
same $\bk$ in the other bundle, and the maps must vary continuously with $\bk$ \cite{milstash}. In our
case, the maps $\phi_\mu$ are those that come from the Syzygy Theorem, viewed as evaluated at the value
$\bk$ in question, so we use the same notation. (What we have done here can be described in the following
way: A vector bundle can be understood as a ``locally-free'' module over the ring of continuous complex
functions on the base space,
which here is the Brillouin zone. Given the exact sequence of modules over $R$ given by the Syzygy Theorem,
we have made a change of rings to obtain the sequence of vector bundles with the same maps $\phi_\mu$.)
However, while the composite maps of $\phi_{\mu+1}$ followed by $\phi_\mu$ do vanish (that is,
$\phi_{\mu+1}\phi_\mu=0$ for all $\mu$), we have not yet shown that the sequence is exact. For a sequence
of maps of vector bundles to be exact, it must be exact when evaluated at any given $\bk$. It follows that
for a sequence of the form shown, for each map the rank [${\rm rank}_{\bf C}\,\phi_\mu(\bk)$] must be the
same everywhere in the base space. That is, all the maps should have full rank at all $\bk$, while we have
seen that for the maps coming from the Syzygy Theorem this might fail.

To prove that the sequence of vector bundles is exact, we use the analyticity of the bundle $E$. As
$\phi_1$ has full rank $m$ at any $\bk_0$, we can assume (by permuting rows) that the first $m$
rows of $\phi_1$ are linearly independent (over $\bf C$) at $\bk_0$. Further, we can assume (by
permuting columns) that the first $m$ columns of this set of rows are linearly independent. We view
this top-left $m\times m$ block of $\phi_1$ as
the $m\times m$ matrix $U$ (which is invertible in the neighborhood of $\bk_0$). Rows of $\phi_1$
(vectors $w$) are determined by their first $m$ entries (the vectors $u$), so we can reduce $\phi_1$
to an $n_1\times m$ matrix $[\phi_1]$ consisting of the first $m$ columns of $\phi_1$. Then we can define
an $(n_1-m)\times m$ matrix $G_\bk$ by the block matrix equation
\be
\left(\begin{array}{c}I_m\\G\end{array}\right)=[\phi_1]U^{-1}.
\ee
Hence $G$ is analytic at $\bk_0$, and we have
\be
\phi_2\left(\begin{array}{c}I_m\\G\end{array}\right)=\phi_2[\phi_1]U^{-1}=0.
\ee
The idea is that $G_\bk$ is analogous to $g_\bk$; it characterizes the image of $\phi_2$ as a bundle
[for each $\bk$, $G_\bk$ is a point in $G(n_1-m,n_1)$]. At this point we essentially repeat for $G_\bk$
the arguments used earlier for $g_\bk$ to conclude that $\phi_2$ has full rank at $\bk_0$, and hence at
all real $\bk$. Working back up the chain with this argument, we establish that all the evaluated maps
$\phi_\mu(\bk)$ have full rank at all real $\bk$.

Given that the sequence of vector bundles is exact, the triviality of the image $E$ (the bundle of
interest) of $\phi_1$ in $E_0$ follows by a standard argument: Because of exactness, the image of $E_d$
is faithful, and hence is a trivial bundle of rank $n_d$ embedded in $E_{d-1}$, and $E_{d-1}$ is also
trivial. It follows that the quotient bundle $E_{d-1}/E_d$ (i.e.\ the quotient in the fiber at each $\bk$)
is also trivial. By exactness, this quotient bundle is isomorphic to the image of $\phi_{d-1}$, and the
argument can be repeated. Hence all the image bundles are trivial, ending with $E$, the image of $\phi_1$.
That is the conclusion of the Theorem. QED. The standard fact used in this step is that any short exact
sequence of vector bundles splits---each bundle is equivalent to a direct sum of the image and the quotient
\cite{milstash}. We combined this with triviality to obtain the result.

This argument can be seen as the natural extension of those for the special cases with additional
hypotheses above, for which the proofs were simply specializations of the general one. Note that we do
not in the end claim to show in the more general cases that there is a set of $m$ generators for the
module $M$ (that is, $m$ polynomial sections) that are linearly
independent at all real $\bk$; the trivialization whose existence is proved by the Theorem may contain
vectors with non-polynomial entries.
The Hilbert Syzygy Theorem was essential to the argument, as it established that the exact sequence of
vector bundles terminates at the left. If we had been working only with analytic functions, as would be
the case for states that are not TNSs, then no such finite exact sequence need exist globally over the
Brillouin zone, though they would exist over smaller open neighborhoods \cite{gunning}. It is only in the
case of polynomial sections and an analytic bundle that we obtain such a general result. The Syzygy
Theorem was the key statement that enabled us to argue from the hypothesis involving a local property
(analyticity) to a result involving a global property (topological triviality).

\subsection{Parent Hamiltonian for analytic bundles}

In order to show that the obstruction to finding a gapped parent Hamiltonian is due to the non-triviality
of the filled and empty-band bundles, and does not extend to other examples with trivial bundles, we
should show that when the bundle associated with a TNS is analytic a gapped parent Hamiltonian does exist.
When both bundles are analytic (trivial or not), a projection operator Hamiltonian (projector onto
states in the empty band, minus the projector onto the states in the filled band) has a spectral gap of
$2$ between the filled and empty states, and its matrix elements (when rows and columns are labeled by
positions $x$ as well as by orbital labels) decay exponentially with the distance. For a TNS, we want to
obtain a strictly short-range parent Hamiltonian, so in general a projection operator will not work. We
can state our general result as a Corollary to the proof of the Theorem:\newline
{\bf Corollary}: Under the conditions of the No-Go Theorem, that is when the filled-band bundle is
both polynomial and analytic, a gapped single-particle parent Hamiltonian
exists. \newline
{\bf Proof}: We use the matrix $\phi_1$ from the proof of the Theorem, whose rows are a set of generators
for all solutions to eq.\ (\ref{v=ug}) over the ring $R$. When the filled-band bundle is analytic,
the Proposition implies that at any $\bk$ these generators span the $m$-dimensional fiber of the bundle.
Then, at all $\bk$, $\phi_1^\dagger\phi_1$ is positive semidefinite, its null-space consists of the
empty-band states, and it has rank $m$ (compare with the examples of parent Hamiltonians in the previous
Section). It is noteworthy that, because the number of generators $n_1$ is finite, the entries of this
matrix are finite. Taking a matrix-valued function of $\bk$ constructed similarly for the empty-band
bundle (say $\widetilde{\phi}_1\widetilde{\phi}_1^\dagger$), and subtracting $\phi_1^\dagger\phi_1$,
we obtain a single-particle Hamiltonian as a matrix with entries that are TPs. The many-body version of
this Hamiltonian has the
TNS as ground state, and has an energy gap between the filled and empty bands at every $\bk$ in the
Brillouin zone. As the matrix elements are TPs, it is strictly short range in position space, as
required. QED.

\subsection{Remarks}

We emphasize that there are topological phases of free fermions in which the bundles
are topologically non-trivial because of symmetry requirements, but are trivial when viewed as
complex vector bundles ignoring symmetries. Our No-Go Theorem does not exclude the possibility of a TNS for
one of these to have a gapped parent Hamiltonian; an example is the one-dimensional Kitaev chain
\cite{KitaevChain} (there are no non-trivial complex vector bundles in one dimension).
However, we do not know if this is possible in general. In cases in which the bundle is non-trivial
as a complex vector bundle (that is, ignoring symmetry), our result prevents a gapped short-range parent
Hamiltonian from existing. This is the situation in the $p\pm ip$ examples in Section
\ref{sec:tns}.

\subsection{Compactly-supported Wannier-type functions}

In this subsection, we wish to explain how the No-Go Theorem can be recast to give a result about what may
be called compactly-supported Wannier-type functions. Usually, Wannier functions are single-particle
states constructed from states in a single band (by inverse Fourier transform from $\bk$-space) that are
in some sense well localized in position space. Often, what is used is a section of a band bundle (say,
a line bundle for a single band, such as one of the filled bands) that is normalized at each $\bk$. This
together, with its translates in real space, spans all single-particle states in the band. More generally,
one can consider a set of $m$ such sections (say, orthonormal ones) for $m$ bands, or for a bundle of
rank $m$. It has been recognized \cite{wannier} that for a topologically non-trivial band, such a
normalized section cannot
be constructed such that it is continuous in $\bk$ (sometimes this is described as the absence of a
suitable choice of gauge over the Brillouin zone, because at each $\bk$ the choice reduces to the
choice of a phase for the state). The reason is that such a choice amounts to a non-vanishing section of
the bundle, which by definition does not exist for a non-trivial line bundle. (Similar remarks apply for
$m$ bands forming a higher-rank non-trivial bundle.) For a ``section'' that is
not a true section because it fails to be continuous at some points in the Brillouin zone (say, on a
set of measure zero), the states at those points are not in the Fourier decomposition of the Wannier
functions, and the functions will have power-law tails in real space, and not be well localized.

In some applications it is of interest to consider {\em compactly-supported} Wannier-type functions, that
is functions constructed from a single band or set of bands, that have compact support in real space, and
span all single-particle functions in the band (see for example Ref.\ \cite{ozolins}). In $\bk$ space, such
compactly-supported functions correspond to polynomial sections of the bundle associated with
a single band or sets of bands (such as our filled-band bundle; we will continue to use that terminology
here). (Once again, because of the use
of translations in real space, it is sufficient to consider only polynomials in the ring $R$.)
Notice that here we are no longer demanding (ortho-)normalization at each $\bk$, but we retain a weaker
version of that property: in order not to overlook any single-particle states, in either finite or
infinite systems, we should require that in $\bk$ space the polynomial sections span
the $m$-dimensional space of single-particle states at every $\bk$ in the Brillouin zone. To increase
the likelihood
that that is possible, we allow the set to contain more than $m$ polynomial sections. Thus we use the
following {\bf Definition}: a set of compactly-supported Wannier-type functions for the filled bands is
a set of polynomial sections that span the single particle states in the filled-band bundle at every
$\bk$ in the Brillouin zone (notice the definition includes the ``completeness'' property in $\bk$ space.)
One might expect that, while the notion of compact support (or polynomials sections) is more restrictive
than usual, allowing for more than $m$ sections allows each of them to vanish at some $\bk$, and then
the set might get around the discontinuities associated with non-vanishing (or with orthornormal sets
of) sections of non-trivial bundles. Nonetheless, we have the following version of the No-Go Theorem:

{\bf Theorem}: if a set of compactly-supported Wannier-type functions exists for a filled-band bundle,
then the bundle is trivial when viewed as a complex vector bundle.

{\bf Proof}: In this case, we take the given set of sections as generators for a module $M$ over $R$,
without using a system of linear equations to define them. By the Syzygy Theorem, there is a free
resolution over $R$ ending in $M$. The generators over $R$ form the rows of a matrix of polynomials
$\phi_1$, as before. (In the present case, the resolution may have one more term at the left than previously
\cite{eisen}, but this makes no difference.) By hypothesis, the bundle over the Brillouin zone has full
rank at each $\bk$. Then the same argument as in the proof of the No-Go Theorem shows that the generators
of the kernel (over $R$) of $\phi_1$ form a matrix $\phi_2$ which has full rank over $\bf C$ at each $\bk$.
Repeating the argument as before shows that each map in the resolution has full rank over $\bf C$ at all
$\bk$ in the Brillouin zone, and hence the filled-band bundle spanned by the generators (at all $\bk$) is
trivial as a complex vector bundle. QED.

The Corollary in Sec.\ \ref{sec:nogo2} above, together with the use of the Hilbert basis theorem as 
above before the discussion of the Syzygy Theorem, shows that for any analytic polynomial bundle, 
sets of compactly-supported Wannier-type functions exist. Conversely, we also point out that if a 
set of compactly-supported Wannier-type functions exists, then the bundle is analytic. To show this, 
one only has to take, for any $\bk_0$, a subset of $m$ of the sections that are linearly independent 
at $\bk_0$; this can always be done. Those $m$ polynomial sections are analytic at $\bk_0$, and so 
(e.g.\ by considering determinants) are then linearly independent throughout a neighborhood of $\bk_0$; 
this means the bundle is analytic.


\section {Non--free-fermion phases}
\label{sec:nonfree}

In the wavefunction in the free Chern-band example, without
the particle-hole transformation,
the component that has $N_+$ particles of U(1) charge $+1$ (say, those in the 1 orbital),
and $N_-$ of charge $-1$ (say, those in the $2$ orbital) has $N_+=N_-$, and is a determinant that
when all particles are well-separated has the form (ignoring boundary conditions, for simplicity)
\be
\det \left(\frac{1}{z_i-w_j}\right) =
\frac{\prod_{i<j}(z_i-z_j)\prod_{k<l}(w_k-w_l)}{\prod_{m,n}(z_m-w_n)},
\label{cauchy}
\ee
where we use complex coordinates [e.g.\ $z=x+iy$ in place of ${\bf x}=(x,y)$] $z_i$, $i=1$, \ldots,
$N_+$ for charge $+1$, $w_k$, $k=1$, \ldots, $N_-$
for charge $-1$ particles, and the equality of the two sides is just the Cauchy determinant identity.
The norm-square of the state $|\psi_A\rangle$ can then be identified as
the partition function of a Coulomb plasma (see e.g.\ Ref.\ \cite{kosterlitz}), with
fugacity $|\kappa_1|^2$ for either type of particle. We will assume $|\kappa_1|$ is small so that
particles are well separated. Then it is known that this plasma is in a screening phase, which
confirms the topological identification of the state.

Next we will take $Q$ copies of this TNS, and impose the constraint that the number of particles in each
orbital must be either $0$ or $Q$ (one from each copy) \cite{fermionPEPS,CooperBeri}. Then the composite
of $Q$ fermions (with U($1$) charge either $+1$ or $-1$) will be regarded as the physical particle of
the state, and is a boson for
$Q$ even, fermion for $Q$ odd. The resulting wavefunction for these particles is then the $Q$th power of
that for the one-copy case. The important point is that such a ``product'' construction yields a state
that is again a TNS (because the constraints are local), but not Gaussian/free-fermion for $Q>1$.

The norm-square of the product wavefunction is (at long distances) again a plasma, but with exponent
$Q$ times larger.  A renormalization-group (RG) analysis applies to these plasmas \cite{kosterlitz}, and
shows that for asymptotically small
positive fugacity, screening occurs for $Q\leq 2$. For $Q=2$, the screening length is exponentially
large in $1/|\kappa_1|$  as $|\kappa_1|\to0$. In this case, the TNS is in the same topological phase
as the Laughlin $\nu=1/2$ state for bosons, or SU($2$) level $1$, phase.
The state possesses fractional-statistics
vortex excitations (semions) that carry fractional U(1) charge, and there are
two (or more, possibly depending on the parent Hamiltonian) ground states for the system on a finite torus
with periodic boundary conditions. For $Q>2$, the plasma does not screen at small $|\kappa_1|$, but does
at sufficiently large $|\kappa_1|$, and then corresponds to the $1/Q$ Laughlin state. This construction
can be extended to obtain, for example, the non-Abelian SU($2$) level $k>1$ Chern-Simons theory
\cite{witten}, or the Read-Rezayi states for bosons \cite{rr99,cgt}. For
these TNSs, we do not at present have parent Hamiltonians.

\section{Conclusion}

It is possible to construct wavefunctions that are TNSs for phases of matter
that possess chiral gapless edge modes. This may have important implications for
numerical simulation or theoretical analysis of such phases. However, the parent Hamiltonians for the
free-fermion TNSs are gapless, as we proved in our No-Go Theorem; it is not clear if the same will be
true for the non-free-fermion chiral TNSs.



\acknowledgments

We thank B. Bauer, F. Verstraete, and M. Zaletel for asking (independently) the question that motivated
this work, and B. Bradlyn, D. Freed, G. Moore, R. Howe, and Z. Wang for discussions. This work was
supported by a Yale Postdoctoral Prize Fellowship (JD) and by NSF grants nos.\ DMR-1005895 and
DMR-1408916 (NR). We thank the Simons
Center for Geometry and Physics at SUNY Stony Brook, where the work was initiated, for its hospitality
for both authors and its support for NR, and the Aspen Center for Physics for its hospitality and
partial support for NR under NSF Grant No.\ PHY-1066293 in 2013 while the first version of this work was
completed.

\appendix

\section{Transforming a TNS with auxiliary variables on sites to a TNS with auxiliary variables
on edges}
\label{app:peps}

\begin{figure}[h]
\begin{center}
	\includegraphics[width=0.48\textwidth]{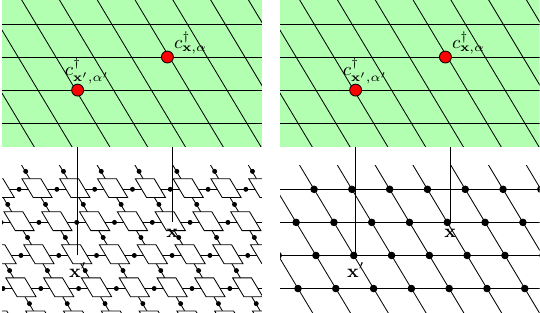}
\end{center}
	\caption{Particles on the square lattice $\mathbb{Z}^2$ (in green) are created by operators
$c^\dagger_{{\bf x},\alpha}$ acting on the vacuum $\kket{0}$, generating a Hilbert (Fock) space of
physical states for each site, $\hilb_{{\bf x}}$. One associates one tensor $T_{{\bf x}} \in
\hilb_{{\bf x}} \otimes V^h  \otimes (V^{h})^* \otimes V^v \otimes (V^{v})^*$ per site ${\bf x}$. The
spaces $V^{h,v}$ are finite-dimensional vector spaces, there is one copy of the space $V^{h,v}$ and
its dual $(V^{h,v})^*$ per (horizontal or vertical) edge of the lattice, represented by solid dots
(left figure). Tracing over all the spaces $V^{h,v}$, namely using the canonical evaluation map
$V \otimes V^* \mapsto \mathbb{C}$ on each edge, we get a state in the many-body Hilbert space
$\bigotimes_{{\bf x} \in \mathbb{Z}^2} \hilb_{{\bf x}}$. In the main text, we used a different version
of a TNS, with auxiliary variables living on the sites, as in the right figure; it can however be
reformulated as a usual TNS.}
	\label{fig:fig2}
\end{figure}

The Gaussian TNS we construct in Section \ref{sec:tns}, with auxiliary variables attached to the physical
sites (illustrated in Fig.~\ref{fig:fig2}-right) can be recast in the form of usual (and in the present
case, still Gaussian) TNS
(Fig.~\ref{fig:fig2}-left). First, let us define what a ``usual TNS'' is. In the case of bosonic
degrees of freedom, a TNS is obtained from a function ${{\bf x}}\mapsto T_{{\bf x}}$ which, to every
site of the square lattice ${{\bf x}} \in \Lambda$, associates a tensor
\begin{equation}
	\label{eq:tensor}
	T_{{\bf x}} \, \in \, \hilb_{{\bf x}} \otimes V^h \otimes (V^h)^* \otimes V^v \otimes (V^v)^*,
\end{equation}
where $\hilb_{{\bf x}}$ is the physical space on the site, and $V^h$, $(V^h)^*$ ($V^v$, $(V^v)^*$) is
the horizontal (vertical) auxiliary space and its dual. Such a tensor is conveniently represented as
follows:
\begin{center}
	\begin{tikzpicture}[scale=1.4]
			\draw (-0.5*3./5.,0.5) -- (0.5*3./5.,-0.5);
			\draw (-0.5,0) -- (0.5,0);
				\filldraw (-0.3,0.5) circle (1.2pt) node[above]{$V^v$};
				\filldraw (0.3,-0.5) circle (1.2pt) node[below]{$(V^v)^*$};
				\filldraw (-0.5,0) circle (1.2pt) node[left]{$(V^h)^*$};
				\filldraw (0.5,0) circle (1.2pt) node[right]{$V^h$};
				\filldraw[fill=white] (0.25-0.3/0.5,0.225) -- ++(0.4,0) -- ++(0.24,-0.4) -- ++(-0.4,0) --
cycle;
		\draw (0,0) node[left]{${\bf x}$} -- (0,0.9) node[above]{$\hilb_{{\bf x}}$};
		\filldraw (0,0.9) circle (1.2pt);
	\end{tikzpicture}
\end{center}
For fermionic degrees of freedom, the construction parallels the one for bosons, but $\hilb_{{\bf x}}$,
$V^h$ and $V^v$ must be $\mathbb{Z}_2$-graded vector spaces, while $(V^h)^*$ and $(V^v)^*$ are their
$\mathbb{Z}_2$-graded duals, and the tensor product $\otimes$ is $\mathbb{Z}_2$-graded (see appendix A
in \cite{ReadSaleur2} for definitions).
It is customary to represent $\mathbb{Z}_2$-graded vector spaces as spaces of polynomials of Grassmann
variables. Indeed, one can view $V^h$ as (isomorphic to) the $\mathbb{Z}_2$-graded vector space
generated by a set of $P$ Grassmann variables living on the edge at position ${\bf x}+{\bf i}/2$:
\begin{equation}
	{\rm span} \{ (\eta^1_{{\bf x}+ {\bf i}/2})^{n_1} \dots (\eta^P_{{\bf x}+ {\bf i}/2})^{n_P} \, | \,
n_1, \dots, n_P = 0,1  \}.
\end{equation}
The $\mathbb{Z}_2$-grading is then the  number (mod $2$) of Grassmann variables in each Grassmann
monomial. Similarly, one can think of $V^v$ as being (isomorphic to) the $\mathbb{Z}_2$-graded vector
space generated by $P$ Grassmann variables on the edge at position ${\bf x}+{\bf j}/2$. The (left) dual
$V^*$ is then generated by another set of $P$ Grassmann variables:
\begin{equation}
	{\rm span} \{ (\eta^{1*}_{{\bf x}+ {\bf j}/2})^{n_1} \dots (\eta^{P*}_{{\bf x}+ {\bf j}/2})^{n_P} \, |
\, n_1, \dots, n_P = 0,1  \}.
\end{equation}
With these notations, the canonical evaluation map is nothing but the Berezin integral:
\begin{equation}
	v^* \otimes u \, \in V^* \otimes V \; \mapsto \; \int [d\eta^1 d\eta^{1*} \dots d\eta^P d\eta^{P*} ]
e^{\sum_p \eta^{p*} \eta^p} \, v^* u .
\end{equation}
The tensor $T_{{\bf x}}$ must have degree $0$, namely it must be a sum of terms with an even total number
of Grassmann variables and physical fermions. Finally, the physical state, which is a state in
$\bigotimes_{{\bf x}\in \mathbb{Z}^2} \hilb_{{\bf x}}$, is defined as ${\rm Tr} \left[
\bigotimes_{{\bf x}} T_{{\bf x}} \right]$, where one traces over all the auxiliary spaces $V^{h,v}$
using the canonical evaluation map (see also Fig.~\ref{fig:fig2}-left).

The Gaussian TNS that we exhibit in Section \ref{sec:tns} have not been expressed in the form of
local tensors
attached to the sites ${\bf x} \in \mathbb{Z}^2$. Instead, they are translation-invariant Gaussian states
of the form
\begin{equation}
	\label{eq:sup_Gaussian}
	 \int [d\xi]  \, \prod_{{\rm edges} \, \left<{\bf x } {\bf y} \right>} e^{\xi^t_{{\bf x}} \cdot A
\cdot \xi_{{\bf y}}} \prod_{{\rm sites}\,{\bf z}}\, e^{\xi^t_{{\bf z}} \cdot B \cdot \xi_{{\bf z}}}
 e^{\xi^t_{{\bf z}} \cdot \kappa \cdot c^\dagger_{{\bf x}}} \kket{0}
\end{equation}
where $\xi_{{\bf x}} = (\xi^1_{{\bf x}},\dots,\xi^P_{{\bf x}})$, $A^{h,v}$ and $B$ are $P
\times P$ matrices, while $\kappa$ is $P\times n$. To recast this expression in the form of a
$\mathbb{Z}_2$-graded TNS, we introduce new
Grassmann variables $\eta_{{\bf x}+ {\bf i}/2}=(\eta^1_{{\bf x} + {\bf i}/2},\dots,\eta^P_{{\bf x} + {\bf
i}/2})$, $\eta_{{\bf x}+ {\bf j}/2}=(\eta^1_{{\bf x} + {\bf j}/2},\dots,\eta^P_{{\bf x} + {\bf j}/2})$,
$\eta^*_{{\bf x}+ {\bf i}/2}=(\eta^{1*}_{{\bf x} + {\bf i}/2},\dots,\eta^{P*}_{{\bf x} + {\bf i}/2})$,
$\eta^*_{{\bf x}+ {\bf j}/2}=(\eta^{1*}_{{\bf x} + {\bf j}/2},\dots,\eta^{P*}_{{\bf x} + {\bf j}/2})$,
and
\begin{eqnarray}
\nonumber	W_{{\bf x}} &=&   \xi^t_{{\bf x}} \cdot  A^h \cdot \eta_{{\bf x}+{\bf i}/2} \, + \,
\xi^t_{{\bf x}} \cdot  A^v \cdot \eta_{{\bf x}+{\bf j}/2} \, + \,  \xi^t_{{\bf x}} \cdot B \cdot \xi_{{\bf
x}} \\
	&&  - \, \xi^t_{{\bf x}}  \cdot \eta^*_{{\bf x}-{\bf i}/2} \, - \, \xi^t_{{\bf x}}  \cdot
\eta^*_{{\bf x}-{\bf j}/2}  ,
\end{eqnarray}
such that the integration over the $\eta$-variables on each edge gives back the exponential weights in
(\ref{eq:sup_Gaussian}). But, instead of tracing out the $\eta$-variables, we now integrate out the
onsite variables $\xi_{{\bf x}}$. This gives a $\mathbb{Z}_2$-graded tensor for each site, as we want:
\begin{equation}
	T_{{\bf x}} \, = \, \int [d \xi_{{\bf x}}] e^{W_{{\bf x}}} e^{\xi_{{\bf x}}^t \cdot \kappa  \cdot
c^\dagger_{{\bf x}}} \, \in  \,\hilb_{{\bf x}} \otimes V^h\otimes (V^h)^* \otimes V^v \otimes (V^v)^* .
\end{equation}

\section{Chern-band example: explicit form}
\label{app:chern}

In Section \ref{sec:tns}, we sketch the construction of a Gaussian TNS which corresponds to a filled
band with
Chern number $1$. Here we give a few more details about this state. We start with a BCS state of the form
$(\ref{eq:sup_Gaussian})$, with $P=4$ Grassmann variables (equivalently, one could write this state with
$P=2$ {\it complex} Grassmann variables). The matrices $A$, $B$ and $\kappa$ are chosen such that:
\begin{subequations}
\begin{eqnarray}
	&& \xi_{{\bf x}+{\bf i}}^t \cdot A^h \cdot \xi_{{\bf x}} \,=\, \\
\nonumber	&&  \left( \begin{array}{cccc} \xi^{1}_{{\bf x}
+{\bf i}} & \xi^{2}_{{\bf x}+{\bf i}}  & \xi^{3}_{{\bf x}+{\bf i}}  & \xi^{4}_{{\bf x}+{\bf i}}   \end{array} \right) \left(\begin{array}{cccc} -i & \lambda &0&0   \\
-\lambda & -i  &0&0 \\
0&0&-i & \lambda   \\
0&0&-\lambda & -i  \end{array} \right) \left( \begin{array}{c} \xi^1_{{\bf x}} \\ \xi^2_{{\bf x}} \\ \xi^3_{{\bf x}} \\ \xi^4_{{\bf x}}
\end{array} \right)   \\
	&& \xi_{{\bf x}+{\bf j}}^t \cdot A^v \cdot \xi_{{\bf x}} \, =  \\
\nonumber	&& \left( \begin{array}{cccc} \xi^1_{{\bf x}
+{\bf j}} & \xi^2_{{\bf x}+{\bf j}}  & \xi^3_{{\bf x}+{\bf j}}  & \xi^4_{{\bf x}+{\bf j}}  \end{array} \right) \left(\begin{array}{cccc} 1 & \lambda   &0 &0 \\
-\lambda & -1 &0&0 \\
0&0& 1 & \lambda   \\
0&0& -\lambda & -1  \end{array} \right) \left( \begin{array}{c} \xi^1_{{\bf x}} \\  \xi^2_{{\bf x}} \\  \xi^3_{{\bf x}} \\ \xi^4_{{\bf x}}
\end{array} \right) \\
&&	\xi_{{\bf x}}^t \cdot B \cdot \xi_{{\bf x}} \, = \\
\nonumber && \left( \begin{array}{cccc} \xi^1_{{\bf x}} &  \xi^2_{{\bf x}} & \xi^3_{{\bf x}} &  \xi^4_{{\bf x}} \end{array} \right) \left(\begin{array}{cccc} 0 & -2\lambda  & 0 &0  \\ 2\lambda  & 0  & 0 &0  \\ 0 & 0 & 0 & -2\lambda \\ 0 & 0 &  2 \lambda & 0 \end{array} \right)
\left( \begin{array}{c} \xi^1_{{\bf x}} \\ \xi^2_{{\bf x}}  \\ \xi^3_{{\bf x}}  \\ \xi^4_{{\bf x}}  \end{array} \right) \\
&&	\xi_{{\bf x}}^t \cdot \kappa  \cdot c^\dagger_{{\bf x}} \, = \, \kappa_1 \, \frac{ \xi^1_{{\bf x}} - i \xi^3_{{\bf x}} }{\sqrt{2}}
 \,c^\dagger_{{\bf x},1} \, +\, \kappa_2 \,  \frac{\xi^{1}_{{\bf x}} + i\xi^{3}_{{\bf x}}  }{\sqrt{2}} \,c^\dagger_{{\bf x},2}  \, .
\end{eqnarray}
\end{subequations}
Here, $\lambda \in \mathbb{R}$, and $\kappa_1$, $\kappa_2 \in \mathbb{C}$ are free parameters. In
momentum space, this state takes the form
\begin{equation}
	\exp \left( \int \frac{d^2 {\bf k}}{(2\pi)^2} \, g_{{\bf k}} \, c^\dagger_{{\bf k},2} c_{-{\bf
k},1}^\dagger  \right) \kket{1},
\end{equation}
where the function $g_{{\bf k}}$ is the following propagator in the auxiliary theory of the Grassmann
variables:
\begin{eqnarray}
	\label{eq:sup_g21}
	\frac{g_{{\bf k}}}{\kappa_1\kappa_2} & = & \frac{\int [\prod d\xi^p] \, e^S \, \frac{ \xi^1_{{\bf k}}
- i \xi^3_{{\bf k}}  }{\sqrt{2}} \frac{ \xi^1_{-{\bf k}} + i \xi^3_{-{\bf k}}  }{\sqrt{2}}}{\int [\prod
d\xi^p ]e^S} \\
\nonumber	& = & \frac{\int [\prod d\xi^p] \, e^S \,  \xi^1_{{\bf k}} \xi^1_{-{\bf k}} }{\int
[\prod d\xi^p ]e^S} \\
\nonumber	&=& \frac{\sin k_x - i \sin k_y}{\left(\sin k_x\right)^2 + \left( \sin k_y \right)^2 +
\lambda^2 \left[ 2 -\cos k_x - \cos k_y  \right]^2}.
\end{eqnarray}
We see that our BCS state depends only on the product $\kappa_1 \kappa_2$, so we can take $\kappa_2 =
\kappa_1$ without loss of generality. Finally, performing the particle-hole transformation mentioned in
the main text, namely $\ket{0} \rightarrow \ket{1,0}$
and $c^\dagger_{{\bf x},1} \rightarrow c_{{\bf x},1}$, we obtain the state
\begin{equation}
	\exp \left( \int \frac{d^2 {\bf k}}{(2\pi)^2} \, g_{{\bf k}} \, c^\dagger_{{\bf k},2} c_{{\bf k},1}
\right) \kket{1,0},
\end{equation}
which corresponds to a filled band; one can check that it has Chern number $1$ as soon as $\lambda \neq 0$.

\section{Alternative way to express a paired state with any $\protect g_\bk=v_\bk/u_\bk$ as
a Gaussian TNS}
\label{app:alt}

For any $g_\bk=v_\bk/u_\bk$, where $u_\bk$ and $v_\bk$ are trigonometric polynomials, we can obtain
the state from a Gaussian TNS that is slightly more general in form. For simplicity, we consider only
the $n=1$ examples, as in Sec.\ \ref{p-ip}. First write $g_\bk$ as $g_\bk=v_\bk v_{-\bk}/(v_{-\bk}u_\bk)$.
We will use a single
Grassmann variable $\xi_{\bf x}$ on each site. For the $\kappa$ terms, we write in $\bk$ space $v_\bk
\xi_{-\bk} c^\dagger_\bk$ for each $\bk$, and for the $A$ and $B$ terms in $\bk$ space
$\xi_{-\bk}v_{-\bk}u_\bk \xi_\bk$. Note that in real space, both kinds of terms are strictly short
range. Then clearly integrating out the $\xi$ variables produces the
desired form. Notice that if $g_\bk$ is odd under $\bk\to-\bk$, then so is $v_{-\bk}u_\bk$, as required
in the Grassmann bilinear form.

This form differs from our original TNS expression in that the coupling of $c_{\bf x}$ to $\xi_{\bf x}$
is no
longer just on site. Moreover, there is now only one variable $\xi$ per site ($P=1$). The first change
compensates the second, so that when both forms exist, the results, including the entanglement spectrum,
must be the same. Then the rank of the entanglement spectrum cannot be read off simply from the number of
$\xi$ variables in the present form. However, due to the TPs (alternatively, polynomials in $R$) in the
coefficients $A$, $B$, $\kappa$, the range of these terms in real space is bounded, and this should
limit the rank of the entanglement spectrum in a similar way. Notice also that many results in the main
text use only the fact that $g_\bk$ is a ratio of TPs, and so hold for the class discussed here.



\end{document}